# Foundation Model for Whole-Heart Segmentation: Leveraging Student-Teacher Learning in Multi-Modal Medical Imaging

Abdul Qayyum[*,1], Moona Mazher[3], Devran Ugurlu[2,1], Jose Alonso Solis Lemus[1], Cristobal Rodero[1], Steven A Niederer[1,2]

[1]National Heart and Lung Institute, Faculty of Medicine, Imperial College London, London, United Kingdom

[2]Turing Research and Innovation Cluster: Digital Twins, The Alan Turing Institute, London, United Kingdom

[3]Hawkes Institute, Department of Computer Science, University College London

*Abstract*— Whole-heart segmentation from CT and MRI scans is crucial for cardiovascular disease analysis, yet existing methods struggle with modality-specific biases and the need for extensive labeled datasets. To address these challenges, we propose a foundation model for whole-heart segmentation using a self-supervised learning (SSL) framework based on a student-teacher architecture. Our model is pretrained on a large, unlabeled dataset of CT and MRI scans, leveraging the xLSTM backbone to capture long-range spatial dependencies and complex anatomical structures in 3D medical images. By incorporating multi-modal pretraining, our approach ensures strong generalization across both CT and MRI modalities, mitigating modality-specific variations and improving segmentation accuracy in diverse clinical settings. The use of large-scale unlabeled data significantly reduces the dependency on manual annotations, enabling robust performance even with limited labeled data. We further introduce an xLSTM-UNet-based architecture for downstream whole-heart segmentation tasks, demonstrating its effectiveness on few-label CT and MRI datasets. Our results validate the robustness and adaptability of the proposed model, highlighting its potential for advancing automated whole-heart segmentation in medical imaging.

## INTRODUCTION

Foundation models have recently emerged as transformative tools in various fields, including language processing and computer vision, due to their remarkable performance [1]. These large-scale deep-learning models trained on vast amounts of unannotated data, serve as versatile foundations for diverse downstream tasks [1-16]. In biomedical image analysis, foundation models have shown significant promise by providing scalable and generalizable frameworks, outperforming traditional methods in critical tasks such as segmentation and classification [3,13,15,16,17]. However, developing these models for biomedical imaging presents unique challenges, primarily due to the limited and specialized datasets available [8,9]. Unlike general domains that benefit from extensively annotated datasets like ImageNet [18] or LAION [19], biomedical imaging datasets are typically small, fragmented, and organ-specific, with scarce annotations [20]. This data limitation highlights the growing need for foundation models specifically tailored to biomedical imaging, designed to leverage larger pre-training datasets and address the field's distinct challenges effectively.

Training deep neural networks traditionally relies on abundant, well-annotated data, which is rarely available in biomedical imaging. To overcome this limitation, alternative strategies such as self-supervised learning (SSL) have emerged [21-23]. SSL enables models to learn meaningful visual representations from unlabeled data through pretext tasks, reducing reliance on annotated datasets [23]. Multi-task learning approaches have also shown potential, aiming to improve performance across various tasks like segmentation, object detection, and classification [3, 4, 20].

Within the field of cardiovascular imaging, cardiac magnetic resonance (CMR) and computed tomography (CT) play a critical role in diagnosing and monitoring heart conditions. Despite the increasing development of foundation models in medical imaging, there remains a significant gap in whole-heart segmentation, particularly in models trained using both CT and MRI data. The existing approach [24] primarily focuses on short-axis segmentation of the heart, often overlooking the need for comprehensive four-chamber whole-heart segmentation. Additionally, previous studies [24] have predominantly relied on MRI alone, excluding CT data, which is crucial for broader clinical applications. Moreover, many of these approaches utilize 2D self-supervised learning (SSL) frameworks such as DINO, which are suboptimal for dense 3D segmentation tasks. These limitations significantly restrict the applicability of existing models in high-fidelity 3D reconstructions, which are essential for computational simulations and mechanical modelling in digital twin applications.

To bridge this gap, we propose a foundation model specifically designed for four-chamber whole-heart segmentation, incorporating both CT and MRI data. Our approach leverages a large-scale, unlabelled dataset for 3D SSL pretraining, ensuring robust feature extraction and generalization across different imaging modalities. The model is then fine-tuned on a limited but carefully curated labeled dataset to achieve precise segmentation performance. By integrating both CT and MRI, our proposed foundation model provides a more comprehensive and clinically relevant solution compared to existing approaches. Furthermore, the ability to generate high-resolution 3D heart meshes enhances its applicability in computational modeling, simulation-based diagnostics, and digital twin frameworks.

The key contributions of this paper are as follows:

1. We developed a self-supervised learning (SSL) framework using a 3D student-teacher architecture trained on a large, unlabeled CT and MRI dataset. Leveraging the xLSTM backbone, our model effectively captures long-range spatial dependencies and complex anatomical structures in 3D medical images.

2. By pretraining on multi-modal datasets, our model generalizes well across both CT and MRI modalities. This cross-modality adaptability mitigates modality-specific biases, ensuring consistent performance across diverse clinical settings and an improvement over existing methods that struggle with such variations. Our approach leverages large-scale unlabeled datasets for pretraining, significantly reducing the need for extensive manual annotations. This enables strong performance even when trained on limited labeled data, addressing a key challenge in medical image segmentation.

3. We validated our proposed method on few-label downstream CT and MRI datasets, demonstrating its effectiveness in generalizing for whole-heart segmentation. Specifically, we introduced an xLSTM-UNet-based model tailored for downstream whole-heart segmentation tasks, further reinforcing the robustness and applicability of our approach.

This work aims to establish a new benchmark in whole-heart segmentation, addressing the limitations of existing models while contributing to advancements in cardiac imaging. The following sections provide a detailed description of the datasets, pretraining methodologies, model architecture, and evaluation metrics used to validate the proposed foundation model.

RELATED WORK

Athira J. Jacob et al. [24] introduced a vision foundation model for CMR assessment, trained in a self-supervised manner on 36 million CMR images and fine-tuned for nine clinical tasks, demonstrating enhanced accuracy, robustness, and few-shot learning capabilities across classification, segmentation, landmark localization, and pathology detection [24]. Matthew Christensen et al. [25] developed EchoCLIP, a vision–language foundation model for echocardiography trained on over one million cardiac ultrasound videos and expert interpretations, demonstrating strong performance in cardiac function assessment, device identification, and clinical transition detection, advancing AI-driven cardiovascular imaging. Kai Zhang et al. [26] introduced BiomedGPT, an open-source, lightweight vision–language foundation model designed for diverse biomedical tasks, achieving state-of-the-art performance in multiple benchmarks while demonstrating strong capabilities in radiology question answering, report generation, and summarisation. George Mathew et al. [27] developed foundation models trained on synchronously captured phonocardiogram (PCG) and ECG data using a self-supervised masked autoencoder framework, enabling superior performance in cardiovascular disease detection despite limited annotated datasets. Yukun Zhou et al. [28] introduced RETFound, a foundation model for retinal images trained on 1.6 million unlabelled images using self-supervised learning, demonstrating superior performance in disease detection and systemic disorder prediction with minimal labeled data. Theodore Zhao et al. [29] introduced BiomedParse, a biomedical foundation model capable of jointly performing segmentation, detection, and recognition across nine imaging modalities, demonstrating superior accuracy and enabling text-driven segmentation for efficient biomedical image analysis. Xiyue Wang et al. [30] introduced CHIEF, a foundation model for histopathology imaging that leverages unsupervised and weakly supervised pretraining to enhance cancer evaluation, demonstrating superior generalizability across diverse populations and digitization protocols. Christian Bluethgen et al. [31] developed a domain-adaptation strategy for large vision–language models, enabling the generation of diverse and medically accurate chest X-ray images from free-form text prompts, overcoming distributional shifts, and augmenting training datasets. Zhi Huang et al. [32] developed OpenPath, a large dataset of over 200,000 pathology images with natural language descriptions, and used it to train the PLIP model, which achieves state-of-the-art performance in pathology image classification and enhances knowledge sharing through image and text-based. Suraj Pa et al. [33] developed a foundation model for cancer imaging biomarker discovery, trained on 11,467 radiographic lesions using self-supervised learning. This model significantly outperformed conventional methods in downstream tasks, especially with limited training data, demonstrating its potential to accelerate the identification of new imaging biomarkers for clinical use.

Ho Hin Lee et al. [34] review the rapid application of the Segment Anything Model (SAM) in biomedical imaging, highlighting its effectiveness in zero-shot learning and its adaptability for diverse medical imaging tasks. While SAM demonstrates state-of-the-art performance in many areas, it still faces challenges in segmenting certain anatomical structures, such as the carotid artery and optic nerve.

Shih-Cheng Huang et al. [35] review the potential of multimodal foundation models, especially Large Vision Language Models (VLMs), to transform healthcare by processing diverse data types and learning from large, unlabeled datasets. They provide a comprehensive analysis of existing research, highlighting the challenges and opportunities for advancing AI systems in medical imaging and offering actionable recommendations for stakeholders across the healthcare sector. Richard J. Chen et al. [36] present UNI, a versatile self-supervised model for computational pathology, trained on more than 100 million images from 20 tissue types. UNI surpasses previous models across various tasks and introduces new features, including resolution-agnostic classification and disease subtyping, providing data-efficient AI solutions for a wide array of challenging clinical applications.

Bastian Wittmann et al. [37] introduce vesselFM, a foundation model specifically developed for 3D blood vessel segmentation. Trained on diverse datasets, vesselFM excels in zero-shot generalization and outperforms existing models across multiple imaging modalities, providing a versatile solution for segmentation in various clinical scenarios.

Zelong Liu et al. [38] introduce VISION-MAE, a foundation model for medical imaging trained on 2.5 million unlabeled images. This model excels in classification and segmentation tasks with high label efficiency, outperforming benchmarks even with limited labeled data, offering a robust solution for diverse.

Yufan He et al. [39] introduce VISTA3D, a unified foundation model for 3D medical imaging that excels in both automatic and interactive segmentation. By integrating a novel 3D superpixel method and leveraging 2D pre-trained backbones, VISTA3D achieves state-of-the-art performance across diverse benchmarks, offering a promising solution for reducing human effort in clinical 3D image analysis.

Suraj Pai et al. [40] introduce CT-FM, a large-scale, pre-trained 3D foundation model designed for radiology tasks, demonstrating superior performance in segmentation, triage, image retrieval, and semantic understanding. By leveraging contrastive learning on 148,000 CT scans, CT-FM excels across diverse radiological tasks and remains robust and interpretable, with open-source access to enhance adaptability in AI solutions for medical imaging.

## MATERIAL & METHODS

### 1. Datasets for SSL and downstream tasks

Our study utilized a diverse range of datasets to develop and evaluate our proposed model. For self-supervised learning (SSL) in whole heart segmentation, we used CT Coronary Angiography (CTCA) [41] images from the Coronary Atlas, ImageCAS (1,000 patients) [42], ImageTBAD (56 CT angiography images for Type-B aortic dissection segmentation) [43], and the TotalSegmentator dataset (1,204 CT scans) [44]. Additionally, the validation datasets from the "Evaluation of Algorithms for Multi-Modality Whole Heart Segmentation" (MMWHS) challenge [45] and the "Whole Heart Segmentation++" (WHS++) challenge [46] were incorporated during SSL, while the training samples from MMWHS and WHS++ were used for downstream tasks. A total of 617 MRI cases were included in this study [47]. The CardioScans dataset [48] consists of 39,200 high-quality CT and MRI DICOM files and 30 anonymized patients, facilitating cardiac imaging research and deep learning applications. Our study utilized 30 cases from this dataset. Additionally, we used the NLST Chest CT dataset [49], which includes 31,801 subjects, as part of the SSL process. The validation datasets from the" Evaluation of Algorithms for Multi-Modality Whole Heart Segmentation" (MMWHS) challenge and the" Whole Heart Segmentation++" (WHS++) challenge were incorporated during SSL, while the training samples from MMWHS and WHS++ were used for downstream tasks. For validation, we utilized the HVSMR-2.0 dataset [50], a 3D cardiovascular MR dataset designed for whole-heart segmentation in congenital heart disease. The dataset provides high-quality cardiac MRI scans that support deep learning applications in medical image analysis. With its comprehensive and well-annotated imaging data, HVSMR-2.0 serves as a reliable benchmark for evaluating our segmentation models, ensuring robust performance assessment in real-world clinical scenarios.

The Medical Image Computing and Computer-Assisted Intervention (MICCAI) challenges have played a significant role in advancing cardiac imaging research, particularly in segmentation and classification tasks. Various publicly available cardiac MRI (CMR) datasets have been used in these challenges, each differing in terms of imaging modality, patient cohort, and anatomical structures of interest. These datasets are essential for developing and evaluating deep-learning models for cardiac image segmentation.

The M&Ms Challenges (2020, 2021) [51-52] provided large multi-center, multi-vendor, and multi-disease datasets for right and left ventricle segmentation, containing 735 subjects and over 12,000 images in cine MRI format. These datasets included short-axis (SAX) and long-axis (LAX) views and were annotated for left ventricle (LV), right ventricle (RV), and blood pool (BP). The Automatic Cardiac Diagnosis Challenge (ACDC) [53] 2017 comprised 100 subjects and 1,902 cine MRI images in SAX view, focusing on the segmentation of LV, RV, and BP. Similarly, the CMRxMotion Challenge (2022) [54] tackled segmentation under respiratory motion artifacts with 160 subjects and 1,730 cine MRI images in SAX view.

The EMIDEC 2020[55] dataset contained 100 subjects and 1,800 late gadolinium enhancement (LGE) MRI images, specifically annotated for myocardial infarction assessment, including left ventricular endocardial and epicardial borders, infarcted regions, and microvascular obstruction (MVO) areas. The Right Ventricle Segmentation Challenge 2012 [56] provided 48 subjects and 768 cine MRI images focusing on endocardial and epicardial RV segmentation. Additionally, the MyoPS 2020 [57] dataset integrated cine, LGE, and T2-weighted MRI images from 75 subjects (560 images) for myocardial pathology segmentation, including scars and edema. The MS-CMRSeg 2019 [59] dataset included 45 subjects and 800 images across multiple MRI sequences such as balanced steady-state free precession (bSSFP), LGE, and cine MRI.

The Sunnybrook Dataset (MICCAI 2009) [60] contained 45 subjects and 768 cine MRI images for LV segmentation, while the Left Ventricle Full Quantification Challenge (2019) [58] featured 100 subjects and 1,120 cine MRI images in SAX view. The Cardiac MRI Reconstruction Challenges (MICCAI 2023, 2024) [61,65] provided cine and T1-weighted MRI datasets with 300–330 subjects and 2,250 images each, with no available labels. Similarly, the 2015 Data Science Bowl Challenge [62] included 1,120 subjects and 22,400 cine MRI images but lacked ground-truth annotations. The MYOSAIQ Challenge (2023) [63] dataset contained 470 subjects and 6,983 LGE MRI images, annotated for LV cavity, healthy myocardium, infarct zones, and MVO.

The MyoPS++ [65] dataset included 250 subjects and 2,700 images across LGE, T2-weighted, and bSSFP sequences, with labels for LV, RV, BP, scars, and edema. The LV Segmentation Challenge [66] and Coronary Artery Disease (CAD) [67] dataset provided cine and LGE MRI images from 200 and 1,204 subjects, respectively, though they lacked labels. The UK Biobank dataset is one of the largest available datasets, comprising 10,000 subjects with SAX and LAX cine MRI images. Of these, 5,000 subjects had labels for LV, RV, left atrium (LA), right atrium (RA), and myocardium, while the remaining 5,000 were unlabeled.

Table 1: Publicly Available CMR Datasets from MICCAI Challenges used in SSL of proposed Foundation 3D-Heart_Seg model

| MICCAI challenges | Number of subjects | Modality | Dim (HxWxDxT) | Views | Target labels |
|---|---|---|---|---|---|
| **Deep Learning Segmentation of the Right Ventricle in Cardiac MRI: The M&Ms Challenge (2021) [51]** | 360 | Cine | 3D | SAX+LAX (4chamber) | LV, RV, BP |
| **Multi-Centre, Multi-Vendor and Multi-Disease Cardiac Segmentation: The M&Ms Challenge. (2020) [52]** | 375 | Cine | 3D+t | SAX | LV, RV, BP |
| **Automatic Cardiac Diagnosis Challenge" dataset (ACDC) 2017 [53]** | 100 | Cine | 3D+t | SAX | LV, RV, BP |
| **The Extreme Cardiac MRI Analysis Challenge under Respiratory Motion (CMRxMotion) 2022 [54]** | 160 | Cine | 3D | SAX | LV, RV, BP |
| **Automatic Evaluation of Myocardial Infarction from Delayed-Enhancement Cardiac MRI (EMIDEC) 2020 [55]** | 100 | LGE | 3D | SAX | left ventricular endocardial and epicardial borders, infarcted areas, and the MVO areas |
| **Right Ventricle Segmentation from Cardiac MRI 212 [56]** | 48 | Cine | 3D | SAX | Endo, Epi |
| **MyoPS: A Benchmark of Myocardial Pathology Segmentation Combining Three-Sequence Cardiac MagneticResonance Images 2020 [57]** | 75 | Cine, LGE | 3D | SAX | LV, BP, RV, scars and edema |
| **Left Ventricle Full Quantification Challenge MICCAI 2019 [58]** | 100 | Cine | 3D | SAX | LV |
| **MS-CMRSeg 2019 Multi-sequence Cardiac MR Segmentation Challenge [59]** | 45 | Cine, LGE, bSSFP | 3D | SAX | LV, BP, RV |
| **The Sunnybrook dataset description MICCAI 2009 [60]** | 45 | Cine | 3D | SAX | LV |
| **Cardiac MRI Reconstruction Challenge MICCAI 2023 [61]** | 300 | Cine, T1 | 3D+t | SAX, LAX | LV, RV, BP |
| **The 2015 Data Science Bowl challenge [62]** | 1120 | Cine | 3D+t | SAX, LAX | No labels available |
| **MYOSAIQ challenge 2023 [63]** | 470 | LGE | 3D | SAX | LV cavity, in the myocardium with healthy tissue, infarct zone, and MVO. |
| **MyoPS++ [64]** | 250 | LGE, T2, bSSFP | 3D | SAX | LV, BP, RV, scars and edema |
| **Cardiac MRI Reconstruction Challenge MICCAI 2024 [65]** | 330 | Cine, T1 | 3D+t | SAX, LAX | No labels available |
| **LV Segmentation Challenge [66]** | 200 | Cine, LGE | 3D+t | SAX, LAX | No labels available |
| **Coronary artery disease (CAD) [67]** | 1204 | Cine, LGE | 3D+t | SAX, LAX | No labels available |
| **UKBioBank Dataset** | 10000 | Cine | 3D+t | SAX, LAX | LV, RV, LA, RA, Myo (5000 labels), 5000 No labels |
| **Total** | 14048 | -- | -- | -- | -- |

For our study, we utilized unlabeled datasets in self-supervised learning (SSL) to pre-train models and then fine-tuned them using labeled datasets for segmentation tasks. The combination of diverse labeled and unlabeled datasets enabled robust feature learning and improved generalization for cardiac image analysis. These datasets collectively offer a comprehensive resource for developing and evaluating AI-driven segmentation models in cardiac imaging research. The details of each dataset are available in Table 1.

**2. Proposed Framework for Whole Heart Segmentation**

Figure 1 presents the overall workflow of the proposed model for whole heart and brain lesion segmentation. The framework comprises four primary stages:

*a)* DATA COLLECTION AND PREPROCESSING

A variety of datasets, including whole heart CT and MRI imaging datasets were curated and preprocessed. The dataset distribution followed three phases: 1. pretraining on large, unlabeled (Cardiac: CT/MRI) datasets for general feature learning, 2. fine-tuning with labeled datasets for the heart (HVSMR-2.0, MMWHS-CT, WHS++CT, MMWHS-MRI, WHS++MRI) segmentation These datasets were split into 80% for training and 20% for testing. 3. Finally, in the testing phase, we evaluated the model on the remaining 20% of labeled data for both heart and brain segmentation tasks to assess its performance after pretraining and fine-tuning. The dataset-splitting strategy is shown in Figure 2.

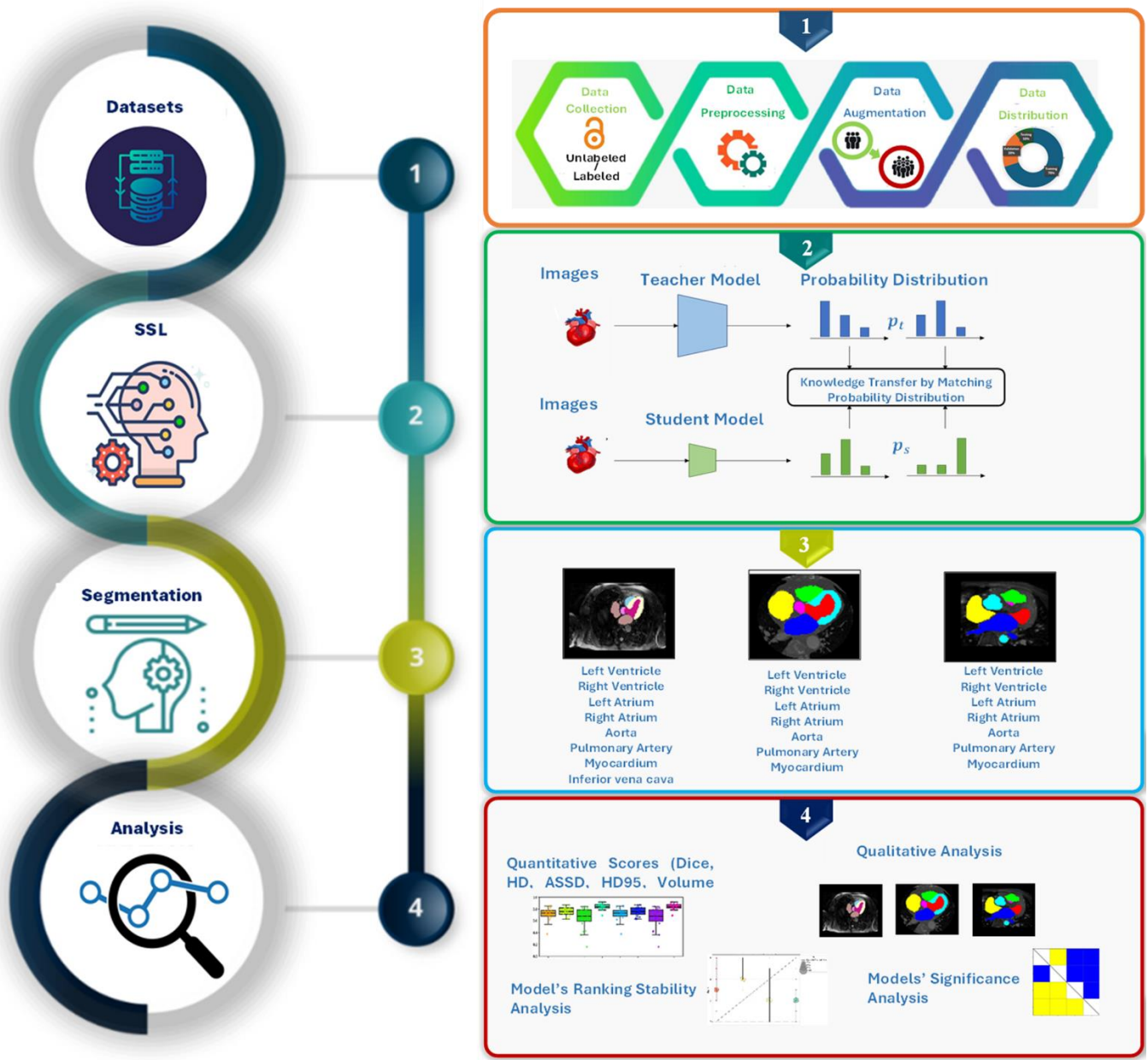

Figure 1. Overall framework encompassing the dataset, self-supervised learning (SSL), and the downstream 3D-Heart_Seg foundation model framework.

*b) INTRODUCTION OF A 3D STUDENT-TEACHER MODEL*

A 3D student-teacher model was designed, inspired by the 2D DinoV2 framework. This model, built on the xLSTM-UNet architecture, was tailored for the SSL phase. It leverages a teacher-student paradigm, where the teacher encoder's parameters are updated using a momentum-based mechanism derived from the student encoder's updates. This iterative process ensures progressive improvements in the teacher's feature representations as the student learns. There are a lot of state-of-the-art segmentation models published in recent years [68-81]. Most of them we have used in comparison with our proposed foundation models.

*c) SUPERVISED FINE-TUNING FOR SEGMENTATION*

In the third stage, the model was fine-tuned using a limited amount of labeled data for segmentation tasks, including whole heart and stroke lesion segmentation. During this phase the encoder and bottom layers pre-trained during SSL were frozen, while the remaining layers were fine-tuned to optimize segmentation performance for specific applications.

*d) Evaluation and Performance Analysis*

The final stage involved rigorous evaluation of the model's performance through comprehensive analysis. The results were benchmarked against state-of-the-art (SOTA) models, demonstrating the superior accuracy and robustness of the proposed approach in whole heart segmentation tasks.

3. **Overview approach for SSL and downstream tasks**

We propose a novel whole-heart segmentation framework leveraging a self-supervised learning (SSL) approach with a 3D xLSTM-UNet model. Our method is structured into three key stages, each designed to improve the model's generalization capability across different imaging modalities, specifically cardiac CT and MRI. The proposed framework is shown in Figure 2.

*a) Stage 1: Self-Supervised Learning (SSL) Pretraining with UKBB Cine MRI*

In the initial stage, we pre-train the xLSTM encoder in a self-supervised learning (SSL) manner using the UK Biobank (UKBB) cine MRI and MICCAI public available Cine and LGE datasets. Cine MRI provides rich spatial and temporal information, making it an ideal dataset for learning cardiac motion dynamics. The SSL pretraining enables the model to learn general cardiac feature representations without requiring labeled annotations. The 3D xLSTM-UNet model is used, where the xLSTM encoder learns temporal dependencies. This stage focuses on learning fundamental cardiac structures and motion characteristics from MRI sequences.

*b) Stage 2: ReSSL (Refined Self-Supervised Learning) with Large-Scale Unlabeled CT & MRI Data*

To further improve the encoder's robustness across modalities, we introduce a Refined Self-Supervised Learning (reSSL) phase, where the encoder undergoes additional self-supervised training using a large-scale dataset comprising both CT and MRI images. This stage extends the learned feature representations beyond MRI to include CT, ensuring better domain adaptation. The reSSL phase utilizes six different unlabeled CT datasets and three different unlabeled MRI datasets, totaling approximately 35,900 cases. The diversity in the dataset helps in learning domain-invariant features. The encoder, previously trained on UKBB MRI data, is retrained using a mix of CT and MRI data in an SSL fashion, further refining its learned representations. This step ensures that the model is capable of handling variations in contrast, resolution, and anatomical differences between the two imaging modalities. We employ five-fold cross-validation to optimize the training process, ensuring model robustness and preventing overfitting.

*c) Stage 3: Supervised Fine-Tuning with Labeled CT & MRI Data*

The primary goal of this stage is to use labeled whole heart CT and MRI datasets to train the model for accurate segmentation of the four cardiac chambers as Left atrium (LA), Right atrium (RA), Left ventricle (LV), Right ventricle (RV), Aorta and Pulmonary artery. Unlike the previous stages that relied on self-supervised training, this stage uses manually annotated CT and MRI datasets where expert radiologists have delineated the four cardiac chambers. The encoder from the previous SSL and reSSL stages is initialized with its learned weights, serving as a pretrained feature extractor. The decoder is trained in a supervised manner using the labeled data to generate precise cardiac segmentation masks. A segmentation loss function (such as Dice loss, cross-entropy loss, or a combination of both) is used to optimize the model for accurate boundary delineation. Since the labeled dataset consists of both CT and MRI images, the model learns to adapt its segmentation capability across different imaging modalities. Various augmentation techniques (such as random rotations, intensity normalization, and elastic deformations) are applied to enhance the model's robustness against anatomical and scanner variations.

*d) Stage 4: Model Validation on Independent Test Set*

After fine-tuning, the final step is to evaluate the trained model on an independent test set to ensure it generalizes well to unseen CT and MRI images. The Dice Similarity Coefficient (DSC) measures the overlap between predicted and ground-truth segmentation masks. Hausdorff Distance (HD) is used to evaluate boundary accuracy by measuring the maximum deviation between predicted and actual contours. The dataset is split into five subsets, where the model is trained on four and tested on one in an iterative manner. This ensures robustness and prevents overfitting. The test set includes data from different hospitals and scanners to evaluate performance across varying clinical settings. The model's performance is compared against other state-of-the-art segmentation techniques, including UNet, nnUNet, and Transformer-based architectures. By integrating SSL, reSSL, and supervised fine-tuning, we develop a foundation model capable of segmenting the whole heart from both CT and MRI with high accuracy. This approach minimizes reliance on large, labeled datasets while ensuring generalizability across different scanners and institutions, making it suitable for real-world clinical applications.

Our approach enables the segmentation of whole heart structures from CT, MRI, or both, making it highly versatile for different clinical settings. Unlike traditional SSL, our reSSL strategy allows the encoder to undergo an additional round of self-supervised learning with new unlabeled data, improving its ability to generalize across modalities.

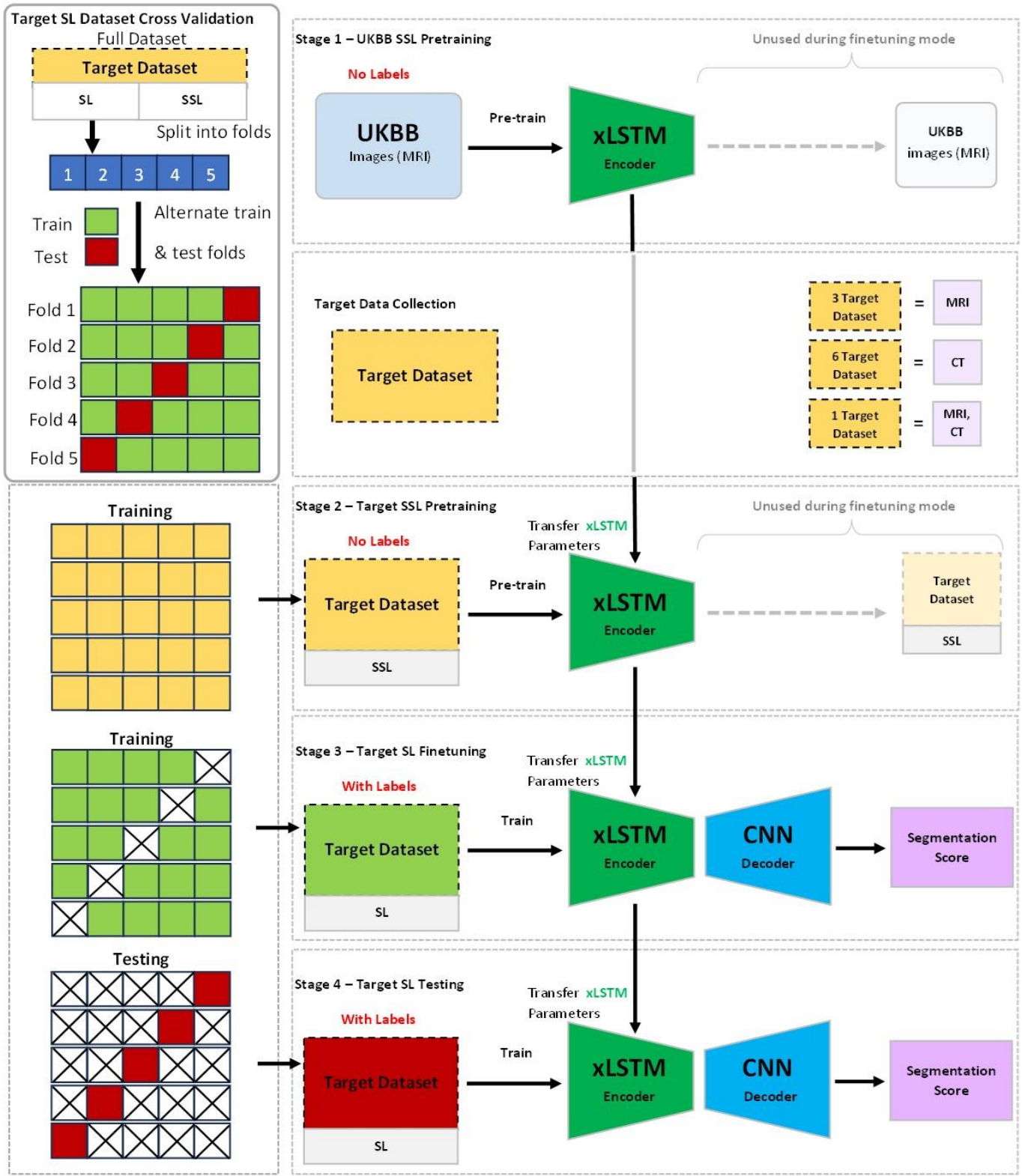

Figure 2. Overview of the proposed framework built upon our Foundation 3D-Heart_Seg model.

By leveraging a vast amount of unlabeled data in two self-supervised learning stages (SSL + reSSL), we significantly reduce the reliance on labeled datasets. The use of five-fold cross-validation ensures that the trained model generalizes well to unseen data. Our proposed 3D xLSTM-UNet-based foundation model effectively segments the four-chamber whole heart from both CT and MRI images. By incorporating self-supervised pretraining, refined SSL training with multimodal unlabeled data, and supervised fine-tuning with limited labeled data, our method achieves strong generalizability across imaging domains. This work represents a step toward a universal deep-learning model for automated whole-heart segmentation in diverse clinical scenarios.

## 4. Methodology

The proposed framework is built on a self-supervised learning (SSL) [70] approach designed to pre-train a 3D Vision-LSTM (xLSTM) integrated UNet model (xLSTM- UNet) [82-83]. The methodology combines advanced deep learning techniques to achieve enhanced performance in 3D medical image segmentation tasks. The main diagram of the proposed SSL model is shown in Figure 3.

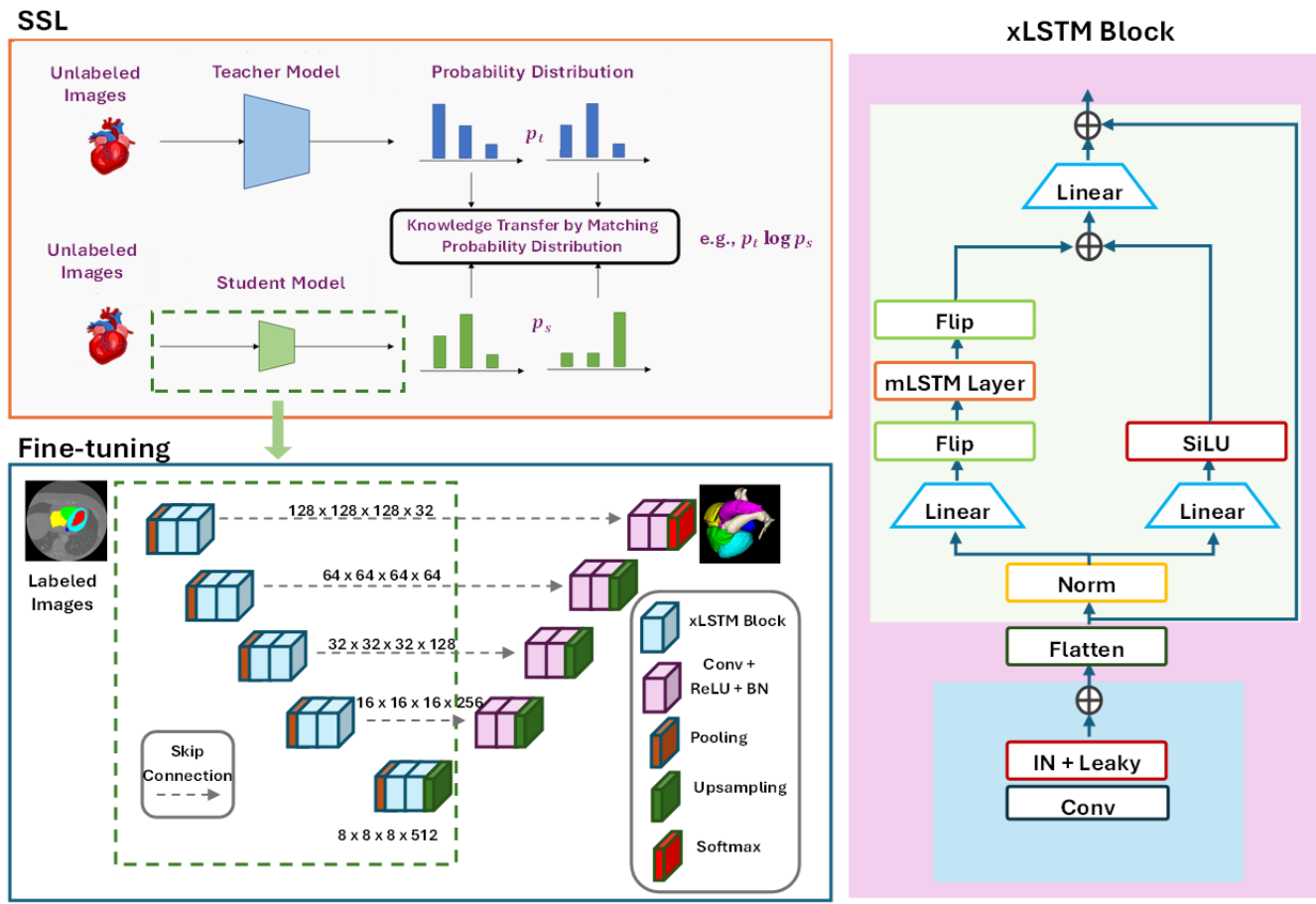

Figure 3. The proposed 3D-Heart_Seg foundation model for whole-heart segmentation.

*a)* Data Augmentation in the Student-Teacher Framework

Robust data augmentation plays a critical role in the SSL pipeline. Techniques such as flipping, scaling, Gaussian noise addition, Gaussian blur, and adjustments to brightness and contrast are applied to create diverse and informative training inputs. Two augmented views of each input image are generated and processed through a Siamese network structure, comprising the student and teacher encoders. The teacher encoder's outputs are refined through centering, sharpening, and normalization via a softmax function, producing supervision signals for the student encoder.

The loss function ensures alignment between the student's outputs and the teacher's processed outputs by minimizing divergence, employing cross-entropy loss and mean squared error (MSE). This alignment facilitates robust feature learning from unlabeled data, enhancing the model's generalization capabilities.

*b)* xLSTM-UNet Architecture

The xLSTM-UNet model integrates Vision-LSTM (xLSTM), an advanced extension of Long Short-Term Memory (LSTM) networks, into the UNet architecture. xLSTM excels at capturing long-range dependencies and contextual information, complementing the UNet's strength in extracting local features through its convolutional encoder-decoder design. The encoder identifies hierarchical features from the input, while the decoder reconstructs these features into detailed segmentation maps, enabling precise and reliable segmentation.

Unlike vanilla LSTMs, which process sequences without specialized optimization for spatial feature learning, xLSTM is designed to capture long-range spatial dependencies across slices more effectively. This is crucial in volumetric medical images, where anatomical structures span multiple slices and require consistent feature propagation. In SSL, xLSTM enhances student-teacher learning by learning more discriminative spatial representations from unlabeled data, ensuring better feature transfer to segmentation tasks with limited labeled data. Furthermore, xLSTM mitigates vanishing gradient issues by incorporating residual connections and adaptive gating mechanisms, allowing for improved gradient flow and better feature retention. It is also more robust to noise and variability in CT/MRI scans, as its advanced recurrent mechanisms suppress irrelevant variations while focusing on essential anatomical structures. Compared to vanilla LSTMs, xLSTM ensures a more efficient and structured information flow, leading to superior segmentation performance with fewer labeled samples. This makes it particularly advantageous for medical imaging applications, where data scarcity and inter-slice dependencies are critical challenges.

*c)* Self-Supervised Pre-Training and Supervised Fine-Tuning

The SSL framework focuses on pre-training the xLSTM-UNet encoder using unlabelled data to capture meaningful spatial and contextual features. Once pre-trained, the encoder is fine-tuned in a supervised manner using labeled datasets, optimizing the decoder to generate accurate segmentation maps. This two-stage process minimizes the reliance on extensive labeled datasets, while the xLSTM module ensures effective learning of global context and long-range dependencies.

To train our proposed 3D Vision-LSTM (xLSTM) model within this framework, we have employed a similar DINOv2 model in a self-supervised learning approach. This methodology allows our model to effectively leverage unlabeled data, improving its ability to learn rich, spatially aware representations that are crucial for downstream tasks such as 3D medical image segmentation. This approach not only maximizes the utility of available data but also enhances the overall performance and robustness of the segmentation model.

The momentum teacher encoder's parameters $\theta_t$ are updated based on the student encoder's parameters $\theta_s$ Using a momentum-based approach:

$$\theta_t = m.\theta_t + (1 - m)\theta_s \quad (1)$$

Where $\theta_t$ are the parameters of the teach encoder, $\theta_s$ are the parameters of the student encoder, $m$ is the momentum coefficient typically a value close to 1.

Let $x$ be the original input image. Two different views of the input, $x_1$ and, $x_2$ are generated using strong data augmentations:

$$x_1 = Augment(x), \ x_2 = Augment(x) \quad (2)$$

Both views are then processed through the student encoder. $f_s$ and teacher encoder $f_t$ to extract feature representations:

$$h_1 = f_s(x_1; \theta_s), \quad h_2 = f_s(x_2; \theta_s) \quad (3)$$
$$h_1' = f_s(x_1; \theta_t), h_2' = f_s(x_2; \theta_t)$$

Where $h_1$ and $h_2$ are the feature representations from the student encoder, $h_1'$ and $h_2'$ are the feature representations from the teacher encoder.

The feature representations $h_1$, $h_2, h_1'$, $h_2'$ are subjected to global average pooling to reduce them into feature vectors:

$$v_1 = GAP(h_1), v_2 = GAP(h_2) \quad (4)$$
$$v_1' = GAP(h_1'), v_2' = GAP(h_2')$$

Where $v_1, v_2, v_1'$ and $v_2'$ are the resulting feature vectors.

$$z_1 = MLP(v_1), z_2 = MLP(v_2) \quad (5)$$
$$z_1' = MLP(v_1'), z_2' = MLP(v_2')$$

After projection, the teacher's output is centered, sharpened, and passed through a softmax function to produce the supervision signal:

$$q_1' = Softmax(\frac{Center(z_1')}{\tau}) \quad (6)$$
$$q_2' = Softmax(\frac{Center(z_2')}{\tau})$$

Where Center (z) subtracts the mean of the vector to have zero mean. $\tau$ is the temperature parameter controlling the sharpness of the distribution. Softmax(z) normalizes the vector into a probability distribution.

The loss function is designed to minimize the divergence between the student's feature vectors and the teacher's processed outputs. A common choice is the cross-entropy loss or mean squared error (MSE) between the student's and teacher's outputs:

$$L = \frac{1}{2}\big(Loss(z_1, q_2') + Loss(z_2, q_1')\big) \quad (7)$$

Where this loss function encourages the student encoder to produce feature representations that align closely with the teacher's outputs, thus enabling effective learning from the unlabeled data.

The cross-entropy loss is

$$L(z, q') = -\sum_{k=1}^{K} q'[k]\log(Softmax(z)[k]) \quad (8)$$

The proposed 3D Vision-LSTM (xLSTM) model is trained using a self-supervised learning (SSL) framework designed to maximize the utility of unlabelled data. Within this framework, the student encoder's parameters are optimized through backpropagation by minimizing the defined loss function L. Simultaneously, the teacher encoder's parameters are updated using a momentum-based mechanism, which incrementally integrates updates from the student encoder to ensure consistent improvement over time. This synergistic combination of self-supervised learning, momentum-driven teacher updates, and comprehensive data augmentation techniques creates a robust and effective pre-training strategy. By leveraging these elements, the model learns meaningful feature representations from unlabelled data, which significantly enhances its ability to deliver accurate and reliable performance in downstream tasks such as 3D medical image segmentation.

## 5. Training and optimization of proposed models

We have developed a self-supervised learning (SSL) framework in PyTorch, tailored for downstream segmentation tasks. We employed the Adam optimizer with a learning rate of 0.00001 for model optimization. The Adam optimizer's adaptive learning rate mechanism helps in handling the sparse gradients often encountered. We selected a very low learning rate to ensure stable convergence, especially when optimizing the student-teacher model with the unlabeled data. To address the downstream segmentation tasks, we combined two loss functions such as cross-entropy loss and Dice loss. Cross-entropy loss is commonly used for pixel-wise classification tasks, as it measures the discrepancy between predicted probabilities and the true class labels. Dice loss, on the other hand, is crucial for segmentation tasks, as it measures the overlap between the predicted segmentation mask and the ground truth, promoting better object boundary delineation. The combination of these two losses ensures that the model not only learns accurate pixel-level classification but also produces segmentation masks with high overlap, enhancing the performance of downstream tasks. The total loss for each patch is computed as the weighted sum of cross-entropy and Dice losses, which helps balance the importance of segmentation performance. During inference, we used a sliding window approach to obtain the final predictions for the entire volume. This technique involves processing the input volume in overlapping patches (windows) to generate predictions at each location. By sliding the window across the volume, we ensure that the model's predictions are computed for all regions, and the overlapping regions help to smooth out edge effects. The predicted patches are then stitched together to form the final, complete segmentation map.

To train the model for both SSL and downstream segmentation tasks, we used a total of 1000 epochs. During each epoch, we applied the optimization strategy and calculated the combined loss during the downstream task. The training process involved a series of iterative updates to both networks. The number of epochs was chosen based on convergence tests, ensuring the model had sufficient time to learn effective representations from the unlabeled data and fine-tune itself for downstream segmentation.

## EXPERIMENTAL RESULTS AND DISCUSSION

Our foundation SSL-based 3D-Heart_Seg model achieved the highest performance across all datasets, demonstrating its superiority over existing state-of-the-art (SOTA) segmentation models. As shown in Table 2, 3D-Heart_Seg consistently outperforms other models, achieving the highest Dice coefficients, such as $0.977 \pm 0.020$ on MMCT++ and $0.931 \pm 0.019$ on MMCT, significantly surpassing other architectures, including Transformer-based, CNN-based, and hybrid models. The results indicate that our self-supervised learning (SSL) approach effectively enhances segmentation performance by leveraging unlabeled data to learn meaningful representations, thereby improving generalization across different datasets. Notably, while other models exhibit performance drops in challenging MRI-based datasets (e.g., HVSMR-2.0), 3D-Heart_Seg maintains a robust performance of $0.771 \pm 0.124$, further proving its adaptability and effectiveness. Compared to Transformer-based models (e.g., VSmTrans) and hybrid architectures (e.g., 3D-xLSTM-UNet and 3D-UMamba), our model demonstrates superior accuracy and consistency, making it the most reliable choice for medical image segmentation. The outstanding performance of 3D-Heart_Seg reinforces the importance of SSL-based techniques in advancing medical image analysis, offering a promising foundation for future large-scale segmentation tasks.

Table 2: Comparative Performance Analysis of Dice Coefficient for the Proposed Foundation Model and State-of-the-Art (SOTA) Segmentation Models

|  | MMCT++ | MMMRI++ | MMCT | MMMRI | HVSMR-2.0 |
|---|---|---|---|---|---|
| 3D-Heart_Seg | **0.977 ± 0.020** | **0.887 ± 0.052** | **0.931 ± 0.019** | **0.871 ± 0.024** | **0.771 ± 0.124** |
| 3D-xLSTM-UNet [82] | 0.929 ± 0.044 | 0.837 ± 0.058 | 0.860 ± 0.066 | 0.818 ± 0.028 | 0.734 ± 0.096 |
| 3D-UMamba [90] | 0.933 ± 0.038 | 0.826 ± 0.088 | 0.886 ± 0.032 | 0.816 ± 0.029 | 0.720 ± 0.117 |
| 3D-nnUNet [91] | 0.909 ± 0.039 | 0.812 ± 0.091 | 0.905 ± 0.008 | 0.824 ± 0.032 | 0.711 ± 0.102 |
| 3D-DensNet [72] | 0.834 ± 0.108 | 0.750 ± 0.084 | 0.841 ± 0.075 | 0.761 ± 0.020 | 0.670 ± 0.137 |
| 3D-ResUNet [72] | 0.865 ± 0.047 | 0.767 ± 0.083 | 0.856 ± 0.025 | 0.739 ± 0.054 | 0.656 ± 0.094 |
| MedNext [92] | 0.890 ± 0.056 | 0.797 ± 0.098 | 0.895 ± 0.018 | 0.803 ± 0.056 | 0.714 ± 0.102 |
| VSmTrans[93] | 0.899 ± 0.061 | 0.802 ± 0.101 | 0.880 ± 0.062 | 0.794 ± 0.064 | 0.694 ± 0.111 |
| LightMUNet [94] | 0.866 ± 0.082 | 0.761 ± 0.075 | 0.872 ± 0.050 | 0.824 ± 0.045 | 0.682 ± 0.116 |
| 3D-UNet [73] | 0.841 ± 0.139 | 0.720 ± 0.187 | 0.805 ± 0.082 | 0.763 ± 0.015 | 0.646 ± 0.144 |
| SAM-Med3D [95] | 0.871 ± 0.059 | 0.773 ± 0.130 | 0.823 ± 0.048 | 0.790 ± 0.025 | 0.674 ± 0.143 |

Our SSL-based 3D-Heart_Seg model achieved the best performance in terms of HD95 (Hausdorff Distance 95%), significantly outperforming state-of-the-art (SOTA) segmentation models across all datasets. As shown in Table 3, our model consistently demonstrates the lowest HD95 values, indicating superior boundary precision and reduced segmentation errors. Specifically, 3D-Heart_Seg achieves an HD95 of 1.251 ± 0.665 on MMCT++ and 4.197 ± 2.518 on MMCT, substantially lower than competing models like 3D-xLSTM-UNet (8.052 ± 11.541 on MMCT++) and 3D-nnUNet (6.834 ± 5.314 on MMCT++). This performance advantage highlights the ability of our model to produce highly accurate and well-aligned segmentation outputs with minimal deviations from ground truth. Furthermore, our model performs exceptionally well on MRI datasets, achieving 6.871 ± 3.264 on MMMRI and 7.757 ± 5.509 on HVSMR-2.0, demonstrating its robustness in handling complex anatomical structures with varying contrasts. In contrast, other models exhibit significantly higher HD95 values, such as 3D-UNet (24.693 ± 4.979 on HVSMR-2.0) and 3D-ResUNet (23.585 ± 20.662 on HVSMR-2.0), indicating poor boundary precision and high segmentation variability.

Table 3: Comparative Performance Analysis of HD95 for the Proposed Foundation Model and State-of-the-Art (SOTA) Segmentation Models

|  | MMCT++ | MMMRI++ | MMCT | MMMRI | HVSMR-2.0 |
|---|---|---|---|---|---|
| 3D-Heart_Seg | **1.251 ± 0.665** | **5.496 ± 3.922** | **4.197 ± 2.518** | **6.871 ± 3.264** | **7.757 ± 5.509** |
| 3D-xLSTM-UNet [83] | 8.052 ± 11.541 | 8.473 ± 5.184 | 10.022 ± 1.336 | 9.029 ± 1.843 | 12.332±10.755 |
| 3D-UMamba [90] | 5.572 ± 7.363 | 7.360 ± 4.453 | 9.707 ± 1.543 | 9.313 ± 3.241 | 10.899 ± 7.881 |
| 3D-nnUNet [91] | 6.834 ± 5.314 | 9.136 ± 4.527 | 9.666 ± 5.038 | 9.538 ± 3.107 | 11.670 ± 6.713 |
| 3D-DensNet [72] | 13.638 ± 9.109 | 13.723 ± 1.870 | 13.821 ± 1.759 | 12.443 ± 8.569 | 17.232 ± 9.876 |
| 3D-ResUNet [72] | 10.704 ± 11.772 | 13.741 ± 2.403 | 12.205 ± 3.237 | 15.264± 2.550 | 23.585±20.662 |
| MedNext [92] | 12.257 ± 14.514 | 11.929 ± 5.453 | 11.280 ± 4.672 | 9.501 ± 4.669 | 13.895±14.117 |
| VSmTrans [93] | 15.358 ± 11.762 | 9.513 ± 7.314 | 9.773 ± 7.451 | 15.081±16.412 | 18.293±24.185 |
| LightMUNet [94] | 10.461 ± 11.725 | 12.669±10.457 | 10.717 ± 1.717 | 11.818 ± 1.973 | 21.244±15.662 |
| 3D-UNet [73] | 17.580 ± 7.555 | 18.547 ± 8.903 | 15.273 ± 5.118 | 16.465 ± 1.946 | 24.693 ± 4.979 |
| SAM-Med3D [95] | 15.239 ± 5.858 | 15.042 ± 6.665 | 14.837 ± 3.113 | 13.994 ± 3.001 | 20.200 ± 7.729 |

The outstanding performance of 3D-Heart_Seg in HD95 evaluation further validates the effectiveness of our self-supervised learning (SSL) approach, which enhances the model's capability to capture fine-grained anatomical details while maintaining structural consistency. By significantly reducing segmentation errors and improving boundary alignment, 3D-Heart_Seg establishes itself as a superior foundation model for medical image segmentation, particularly in complex multi-modality datasets.

Table 4 presents a comparative analysis of the proposed SSL-based foundation model, 3D-Heart_Seg, against several state-of-the-art (SOTA) self-supervised learning (SSL) models in terms of Dice coefficient performance across five benchmark datasets: MMCT++, MMMRI++, MMCT, MMMRI, and HVSMR-2.0. The results demonstrate that 3D-Heart_Seg consistently outperforms all SOTA SSL models across all datasets, highlighting its superior segmentation accuracy.

In the MMCT++ dataset, 3D-Heart_Seg achieves the highest Dice score of 0.977 ± 0.020, significantly surpassing other SSL models such as Voco (0.94 ± 0.123), CADS (0.92 ± 0.089), Hi-End-MAE (0.91 ± 0.123), SwinMM (0.924 ± 0.095), and PCRLv2 (0.913 ± 0.049). This indicates that 3D-Heart_Seg excels in capturing structural details and improving segmentation quality compared to existing models. Similarly, in the MMMRI++ dataset, the foundation model attains 0.887 ± 0.052, outperforming UniMiSS+ (0.839 ± 0.029), SwinSSL (0.811 ± 0.031), and CADS (0.821 ± 0.987).

For the MMCT dataset, the proposed model achieves 0.931 ± 0.019, demonstrating superior performance over SwinSSL (0.871 ± 0.089), CADS (0.880 ± 0.098), and SwinMM (0.911 ± 0.080). Similarly, in the MMMRI dataset, 3D-Heart_Seg attains 0.871 ± 0.024, surpassing Hi-End-MAE (0.818 ± 0.061), UniMiSS+ (0.819 ± 0.011), and PCRLv2 (0.832 ± 0.089). The highest performance gain is observed in the HVSMR-2.0 dataset, where 3D-Heart_Seg achieves 0.771 ± 0.124, outperforming all other models, including PCRLv2 (0.748 ± 0.328), SwinSSL (0.719 ± 0.022), and Hi-End-MAE (0.729 ± 0.111).

Table 4: Performance Comparison of the Proposed SSL Foundation Model with State-of-the-Art SSL Models Based on Dice Coefficients

|  | MMCT++ | MMMRI++ | MMCT | MMMRI | HVSMR-2.0 |
|---|---|---|---|---|---|
| 3D-Heart_Seg | **0.977 ± 0.020** | **0.887 ± 0.052** | **0.931 ± 0.019** | **0.871 ± 0.024** | **0.771 ± 0.124** |
| Voco [85] | 0.94 ± 0.123 | 0.846 ± 0.765 | 0.891 ± 0.089 | 0.837 ± 0.011 | 0.741 ± 0.011 |
| CADS [86] | 0.92 ± 0.089 | 0.821 ± 0.987 | 0.880 ± 0.098 | 0.827 ± 0.011 | 0.731 ± 0.239 |
| Hi-End-MAE [87] | 0.91 ± 0.123 | 0.839 ± 0.187 | 0.916 ± 0.079 | 0.818 ± 0.061 | 0.729 ± 0.111 |
| SwinMM [68] | 0.924± 0.095 | 0.835 ± 0.091 | 0.911 ± 0.080 | 0.829 ± 0.087 | 0.729 ± 0.897 |
| SwinSSL [69] | 0.904 ± 0.698 | 0.811 ± 0.031 | 0.871 ± 0.089 | 0.839 ± 0.071 | 0.719 ± 0.022 |
| UniMiSS+ [88] | 0.936 ± 0.768 | 0.839 ± 0.029 | 0.901 ± 0.098 | 0.819 ± 0.011 | 0.739 ± 0.198 |
| PCRLv2 [89] | 0.913 ± 0.049 | 0.829 ± 0.081 | 0.907 ± 0.092 | 0.832 ± 0.089 | 0.748 ± 0.328 |

Overall, these results demonstrate the effectiveness of 3D-Heart_Seg, which consistently outperforms SOTA SSL models by leveraging advanced self-supervised learning strategies. The model exhibits higher segmentation accuracy across multiple datasets, making it a robust and generalizable approach for medical image segmentation. The significant performance improvements suggest that 3D-Heart_Seg effectively learns spatial representations and structural details, ensuring accurate and reliable segmentation across diverse imaging datasets.

Table 5 provides a comparative analysis of the Hausdorff Distance 95 (HD95) performance of the proposed SSL-based foundation model, 3D-Heart_Seg, against various state-of-the-art (SOTA) self-supervised learning (SSL) models across five benchmark datasets: MMCT++, MMMRI++, MMCT, MMMRI, and HVSMR-2.0. The HD95 metric measures the worst-case segmentation error, with lower values indicating more accurate and reliable segmentation. The results demonstrate that 3D-Heart_Seg consistently achieves the lowest HD95 values across all datasets, signifying superior segmentation boundary precision compared to SOTA SSL models. For the MMCT++ dataset, 3D-Heart_Seg achieves the lowest HD95 value of 1.251 ± 0.665, significantly outperforming other SSL models such as Voco (4.367 ± 2.112), CADS (6.116 ± 1.456), Hi-End-MAE (7.198 ± 1.456), SwinMM (6.897 ± 2.341), and PCRLv2 (7.897 ± 10.231). Similarly, in the MMMRI++ dataset, 3D-Heart_Seg attains an HD95 value of 5.496 ± 3.922, outperforming models like SwinSSL (10.189 ± 1.615), Hi-End-MAE (8.937 ± 5.201), and UniMiSS+ (8.235 ± 2.871), demonstrating its ability to achieve better boundary accuracy.

Table 5: Comparative Analysis of the Proposed SSL-Based Foundation Model and SOTA SSL Models Using HD95 Metric

|  | MMCT++ | MMMRI++ | MMCT | MMMRI | HVSMR-2.0 |
| --- | --- | --- | --- | --- | --- |
| 3D-Heart_Seg | **1.251 ± 0.665** | **5.496 ± 3.922** | **4.197 ± 2.518** | **6.871 ± 3.264** | **7.757 ± 5.509** |
| Voco [85] | 4.367 ± 2.112 | 6.897 ± 2.198 | 6.189 ± 2.891 | 8.987 ± 2.198 | 10.198±3.678 |
| CADS [86] | 6.116 ± 1.456 | 8.184 ± 3.791 | 7.871 ± 2.556 | 11.109 ± 1.327 | 11.023 ± 3.876 |
| Hi-End-MAE [87] | 7.198 ± 1.456 | 8.937 ± 5.201 | 8.231 ± 2.986 | 10.619 ± 4.109 | 12.023 ± 4.798 |
| SwinMM [68] | 6.897 ± 2.341 | 9.115 ± 4.181 | 9.162 ± 2.101 | 9.976 ± 2.311 | 11.897 ± 2.678 |
| SwinSSL [69] | 9.897 ± 7.289 | 10.189 ± 1.615 | 9.987 ± 3.098 | 10.990± 2.786 | 13.678±4.233 |
| UniMiSS+ [88] | 7.631 ± 4.456 | 8.235 ± 2.871 | 8.191 ± 2.158 | 8.894 ± 2.498 | 11.876±4.678 |
| PCRLv2 [89] | 7.897 ± 10.231 | 8.019 ± 2.891 | 7.938 ± 3.169 | 9.198±2.234 | 9.987±6.876 |

In the MMCT dataset, 3D-Heart_Seg achieves an HD95 of 4.197 ± 2.518, indicating a notable improvement over other models such as SwinMM (9.162 ± 2.101), SwinSSL (9.987 ± 3.098), and CADS (7.871 ± 2.556). Similarly, for the MMMRI dataset, 3D-Heart_Seg attains an HD95 of 6.871 ± 3.264, significantly outperforming Voco (8.987 ± 2.198), CADS (11.109 ± 1.327), and PCRLv2 (9.198 ± 2.234). The HVSMR-2.0 dataset showcases the largest performance gap, where 3D-Heart_Seg achieves the lowest HD95 of 7.757 ± 5.509, significantly lower than SwinSSL (13.678 ± 4.233), Hi-End-MAE (12.023 ± 4.798), CADS (11.023 ± 3.876), and SwinMM (11.897 ± 2.678). The lower HD95 values across all datasets indicate that 3D-Heart_Seg produces more precise segmentations with minimal boundary errors, making it a highly effective model for medical image segmentation.

Overall, these results underscore the superiority of 3D-Heart_Seg over existing SSL-based segmentation models by effectively reducing segmentation errors and improving boundary accuracy. The significant reduction in HD95 values demonstrates the model's capability to maintain structural integrity, making it an optimal choice for medical image segmentation tasks requiring high precision and robustness.

We conducted a comprehensive evaluation of our proposed Heart_seg3d foundation model across multiple datasets, including three widely recognized whole-heart segmentation datasets: MMWHS, WHS++, and HVSMR-2.0.

Our experimental results, as depicted in Figures 4-7 consistently demonstrate that Heart_seg3d surpasses SOTA models in both Dice score and Hausdorff Distance 95% (HD95) across various cardiac imaging datasets, including both CT and MRI modalities. The model's superior performance is primarily attributed to the integration of self-supervised learning (SSL), which efficiently leverages unlabelled MRI and CT datasets, thereby enhancing generalization across complex segmentation tasks.

We evaluated the Blob Plot using the Dice score of the proposed 3D-Heart-Seg foundation model and compared its performance with state-of-the-art (SOTA) segmentation models using the HVSMR-2.0 dataset as shown in Figure.6. The analysis reveals distinct performance patterns among the evaluated methods. 3D-Heart-Seg emerges as the most reliable model, achieving the lowest median rank (around Rank 2-3) with minimal variability, indicating consistent performance across bootstrap samples. 3D-XLSTM-UNet also performs well, though with slightly higher variability. Moderate-performing models, including 3D-mamba, 3D-nnUNet, 3D-DenseUNet, and 3D-ResUNet, exhibit wider error bars and median ranks around 5-6, suggesting performance fluctuations across different bootstrap samples. In contrast, MedNext and VsmTrans rank slightly lower, stabilizing around Rank 6-7. The lowest-performing models, such as LightMUNet, 3D-UNet, and SAM-Med3D, show high median ranks (~8-10) with significant variability, indicating instability and sensitivity to data variations. These findings suggest that models with tighter error bars, such as 3D-Heart-Seg and 3D-XLSTM-UNet, offer the most consistent performance. We observe similar performance trends is evaluated on the MMMRI, and MMMRI++ datasets. These results further validate the robustness and generalizability of our approach across different datasets and imaging modalities.

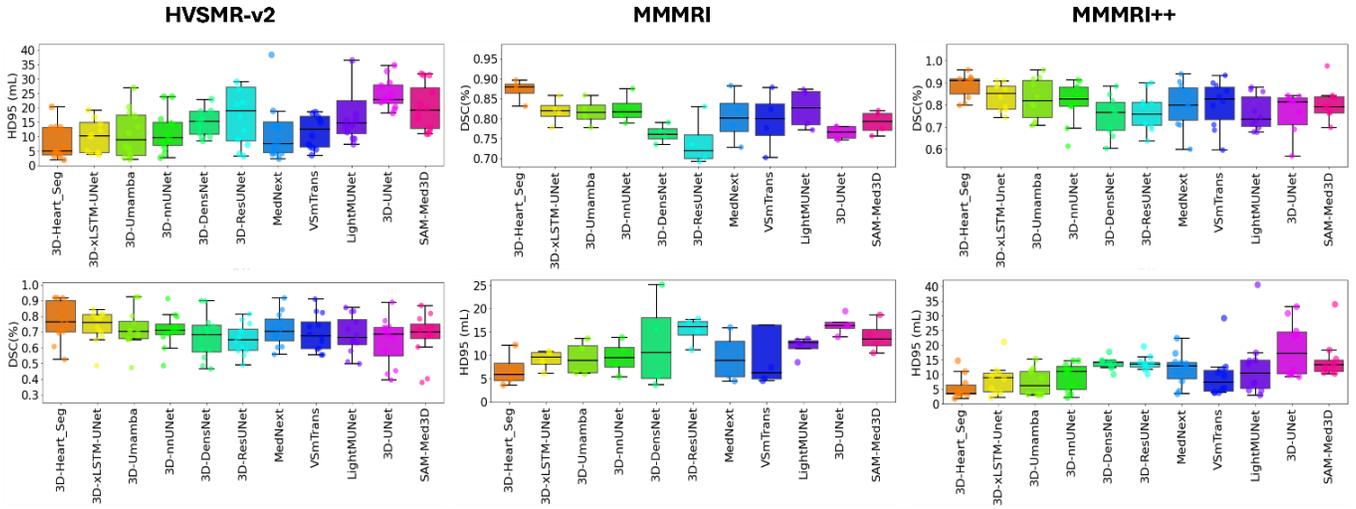

Figure.4. Performance analysis of foundation model and SOTA model using Dice HD95 for HVSMR-v2, MMMRI, MMMRI++ datasets

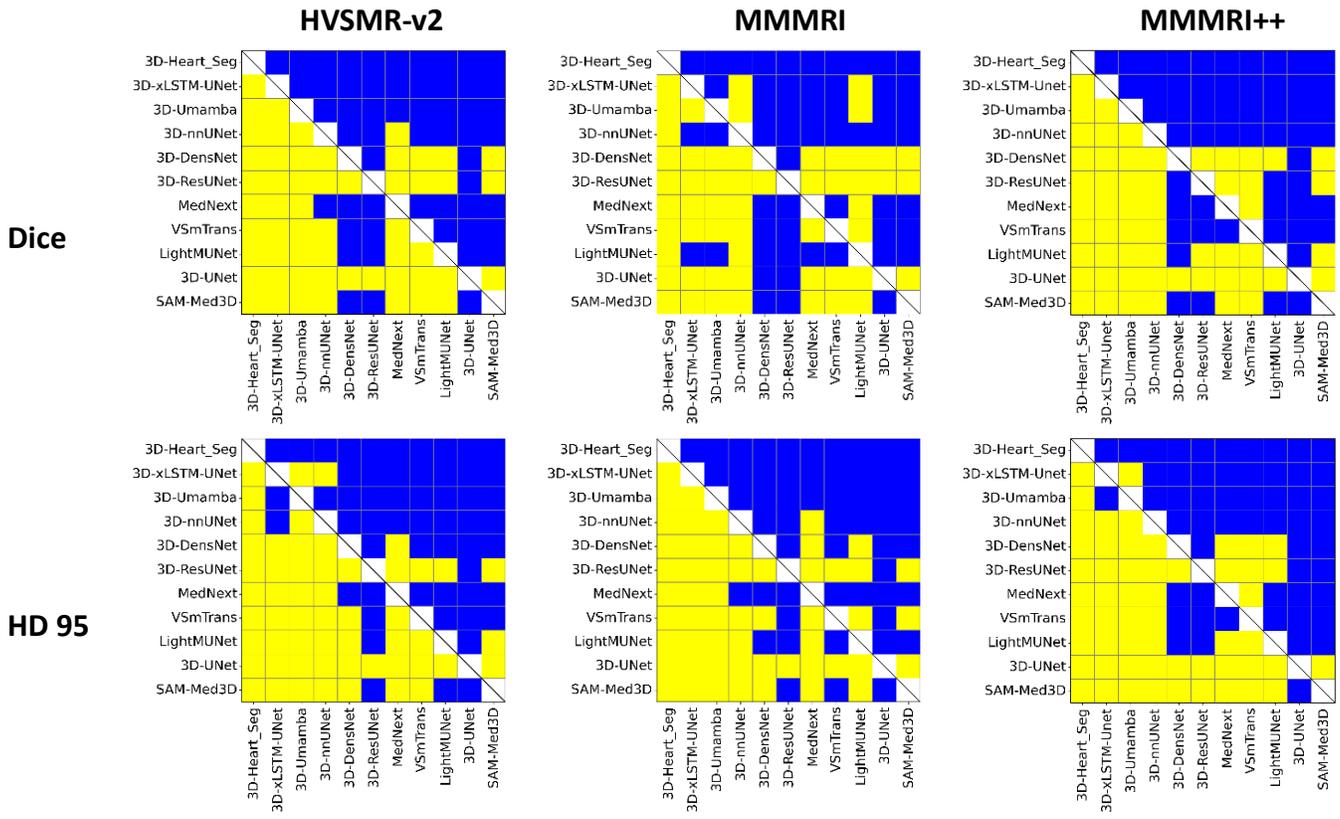

Figure.5. Performance analysis of foundation model and SOTA model using Dice HD95 for HVSMR-v2, MMMRI, MMMRI++ datasets.

Similarly, we evaluated model performance using HD95 (Hausdorff Distance 95th percentile), with results presented in Figures 4-6. Our proposed 3D-Heart-Seg and 3D-xLSTM-UNet consistently achieve the most stable performance across datasets. In contrast, models with higher median ranks exhibit greater variability in HD95, indicating increased sensitivity to data variations. This reinforces the robustness of our approach, as lower HD95 values reflect more precise and reliable segmentation, particularly in challenging anatomical regions.

Our proposed 3D-Heart-Seg foundation model achieved the highest performance based on both Dice score and HD95, as demonstrated in the Matrix Plot analysis. The pairwise ranking comparison reveals that 3D-Heart-Seg consistently outperforms most other segmentation models, with a dominant presence of yellow cells in the top row, indicating superior ranking performance. Similarly, 3D-XLSTM-UNet also performs well but ranks slightly lower than 3D-Heart-Seg in some cases. In contrast, lower-performing models, such as 3D-UNet and SAM-Med3D, exhibit mostly blue cells, reflecting their weaker performance. Mid-tier models, including MedNext, VsmTrans, and 3D-ResUNet, display mixed rankings, suggesting variability in their performance across different scenarios. These results confirm that 3D-Heart-Seg is the most robust and reliable segmentation model, demonstrating both high accuracy (Dice) and precise boundary delineation (HD95) compared to state-of-the-art methods.

Statistical significance maps presented in Figure 5,7 provide further insights into the performance advantages of Heart_seg3d. The model exhibits widespread yellow regions, representing statistically significant improvements across all evaluated metrics compared to competing models. In contrast, SOTA models show fewer yellow regions, indicating weaker performance, while 3D-UNet is marked by blue regions, signifying considerably lower performance levels. The multi-layer SSL pre-training strategy employed in Heart_seg3d plays a pivotal role in achieving significantly higher Dice scores and lower HD95 values compared to existing SOTA models. This approach enhances the model's ability to capture richer hierarchical feature representations, contributing to improved segmentation performance. Heart_seg3d's architecture balances fine-grained local detail with a holistic anatomical understanding, enabling precise segmentation of complex structures. Heart_seg3d demonstrates superior segmentation accuracy on labeled datasets, including, HVSMR-v2 whole-heart MMMRI, MMMRI++ (Figure.5) and MMCT, MMCT++(Figure.7).

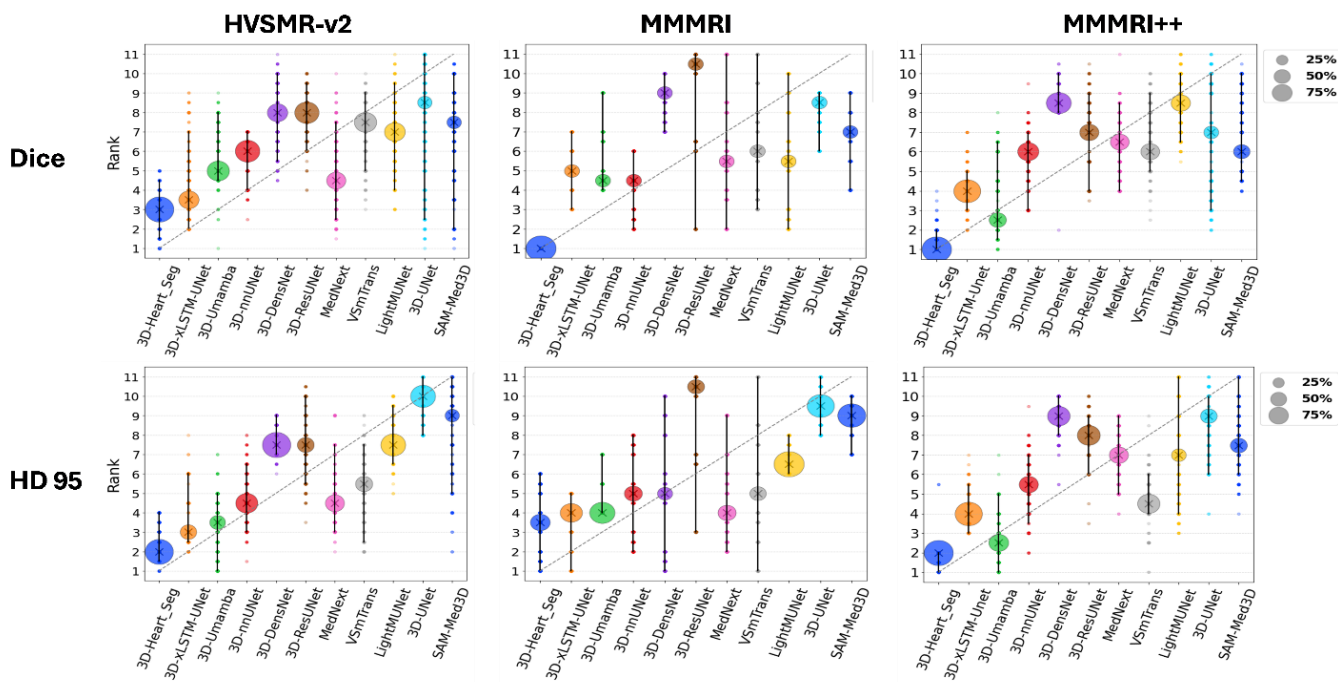

Figure.6. Performance analysis of foundation model and SOTA model using Dice HD95 for HVSMR-v2, MMMRI, MMMRI++ datasets

As illustrated in Figures 8-9 our proposed Heart_Seg3D model demonstrates better performance in segmenting key cardiac structures, including the left ventricle (LV), right ventricle (RV), left atrium (LA), right atrium (RA), aorta (AO), and pulmonary artery (PA). The model consistently achieves higher Dice scores across these structures, showcasing its ability to accurately delineate both large and small anatomical components, even in challenging segmentation scenarios. On the HVSMR-2.0 dataset as shown in Figure.10, Heart_Seg3D outperforms competing models in segmenting the ventricles and atria, achieving higher Dice scores and lower HD95 values. The aorta (AO) and pulmonary artery (PA), which are often more difficult to segment due to their complex branching structures and variability across patients, are also well delineated by our model. This highlights the effectiveness of our self-supervised pretraining strategy, which enables robust generalization across varying anatomical structures.

Similarly, we evaluated model performance using HD95 (Hausdorff Distance 95th percentile), with results presented in Figures 4-6. Our proposed 3D-Heart-Seg and 3D-xLSTM-UNet consistently achieve the most stable performance across datasets. In contrast, models with higher median ranks exhibit greater variability in HD95, indicating increased sensitivity to data variations. This reinforces the robustness of our approach, as lower HD95 values reflect more precise and reliable segmentation, particularly in challenging anatomical regions.

Our proposed 3D-Heart-Seg foundation model achieved the highest performance based on both Dice score and HD95, as demonstrated in the Matrix Plot analysis. The pairwise ranking comparison reveals that 3D-Heart-Seg consistently outperforms most other segmentation models, with a dominant presence of yellow cells in the top row, indicating superior ranking performance. Similarly, 3D-XLSTM-UNet also performs well but ranks slightly lower than 3D-Heart-Seg in some cases. In contrast, lower-performing models, such as 3D-UNet and SAM-Med3D, exhibit mostly blue cells, reflecting their weaker performance. Mid-tier models, including MedNext, VsmTrans, and 3D-ResUNet, display mixed rankings, suggesting variability in their performance across different scenarios. These results confirm that 3D-Heart-Seg is the most robust and reliable segmentation model, demonstrating both high accuracy (Dice) and precise boundary delineation (HD95) compared to state-of-the-art methods.

Statistical significance maps presented in Figure 5,7 provide further insights into the performance advantages of Heart_seg3d. The model exhibits widespread yellow regions, representing statistically significant improvements across all evaluated metrics compared to competing models. In contrast, SOTA models show fewer yellow regions, indicating weaker performance, while 3D-UNet is marked by blue regions, signifying considerably lower performance levels. The multi-layer SSL pre-training strategy employed in Heart_seg3d plays a pivotal role in achieving significantly higher Dice scores and lower HD95 values compared to existing SOTA models. This approach enhances the model's ability to capture richer hierarchical feature representations, contributing to improved segmentation performance. Heart_seg3d's architecture balances fine-grained local detail with a holistic anatomical understanding, enabling precise segmentation of complex structures. Heart_seg3d demonstrates superior segmentation accuracy on labeled datasets, including, HVSMR-v2 whole-heart MMMRI, MMMRI++ (Figure.5) and MMCT, MMCT++(Figure.7).

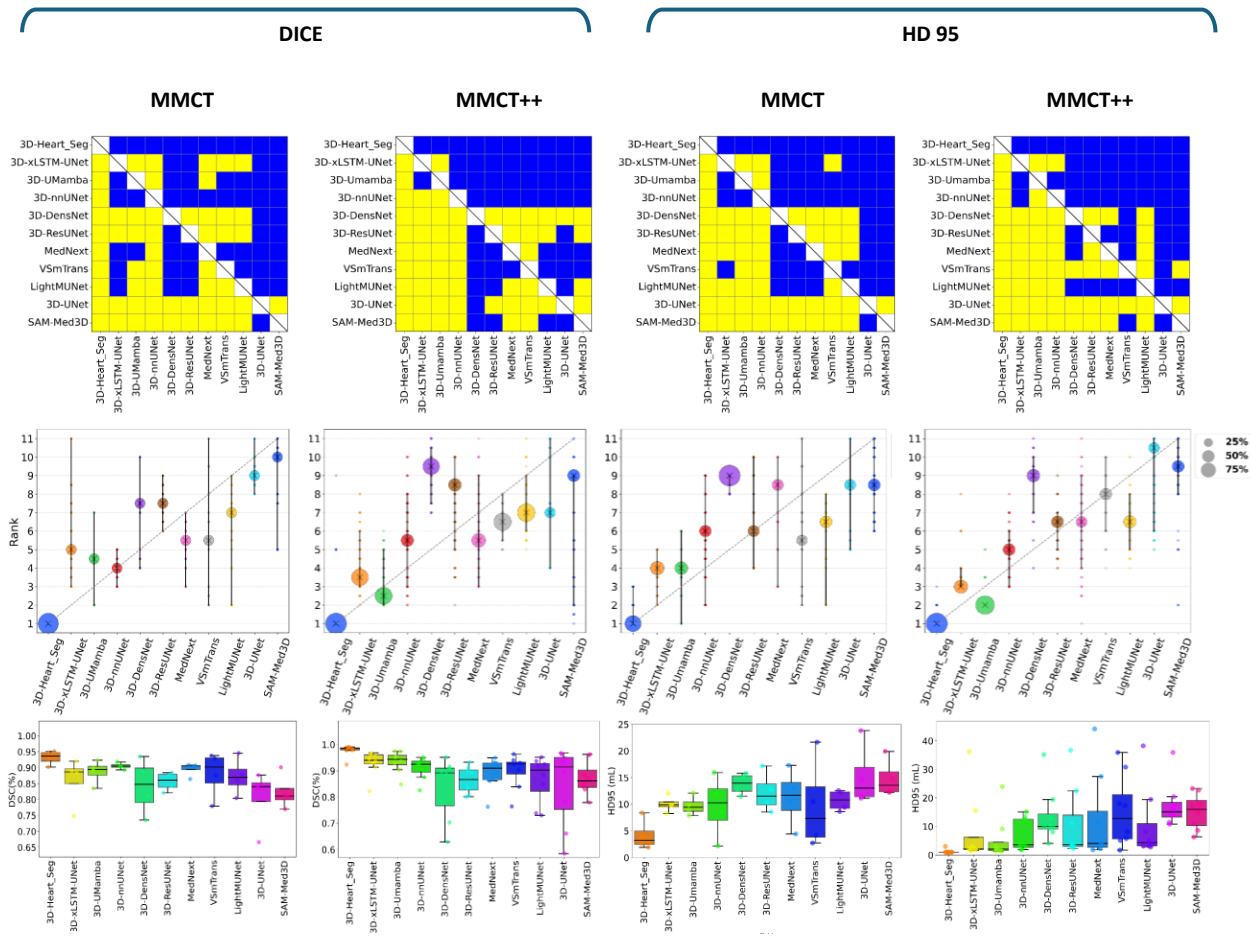

Figure.7. Performance analysis of foundation model and SOTA model using Dice HD95 for MMCT, MMCT++ datasets

As illustrated in Figures 8-9 our proposed Heart_Seg3D model demonstrates better performance in segmenting key cardiac structures, including the left ventricle (LV), right ventricle (RV), left atrium (LA), right atrium (RA), aorta (AO), and pulmonary artery (PA). The model consistently achieves higher Dice scores across these structures, showcasing its ability to accurately delineate both large and small anatomical components, even in challenging segmentation scenarios. On the HVSMR-2.0 dataset as shown in Figure.10, Heart_Seg3D outperforms competing models in segmenting the ventricles and atria, achieving higher Dice scores and lower HD95 values. The aorta (AO) and pulmonary artery (PA), which are often more difficult to segment due to their complex branching structures and variability across patients, are also well delineated by our model. This highlights the effectiveness of our self-supervised pretraining strategy, which enables robust generalization across varying anatomical structures.

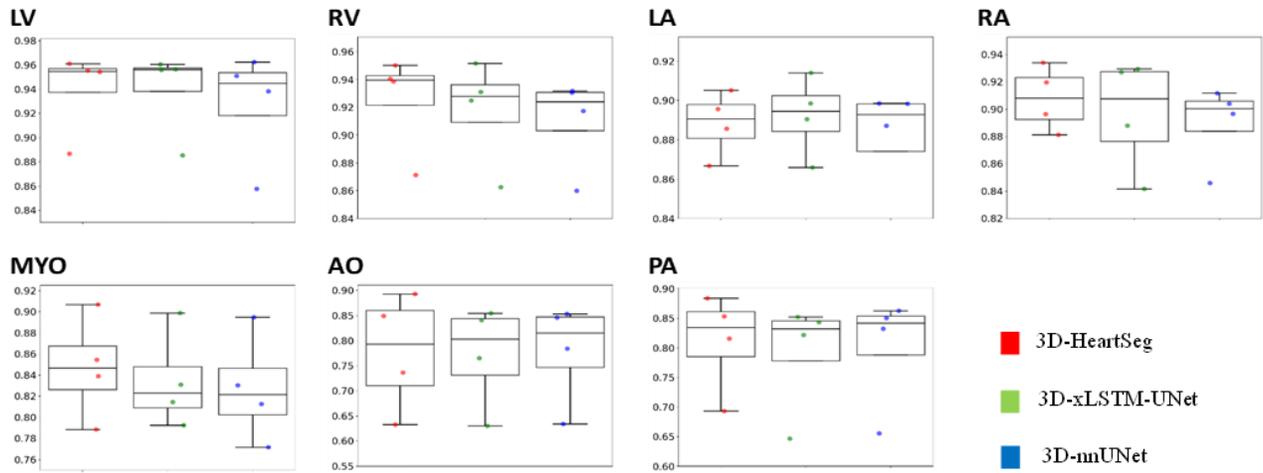

Figure 8. Performance analysis of each class in MMMRI dataset.

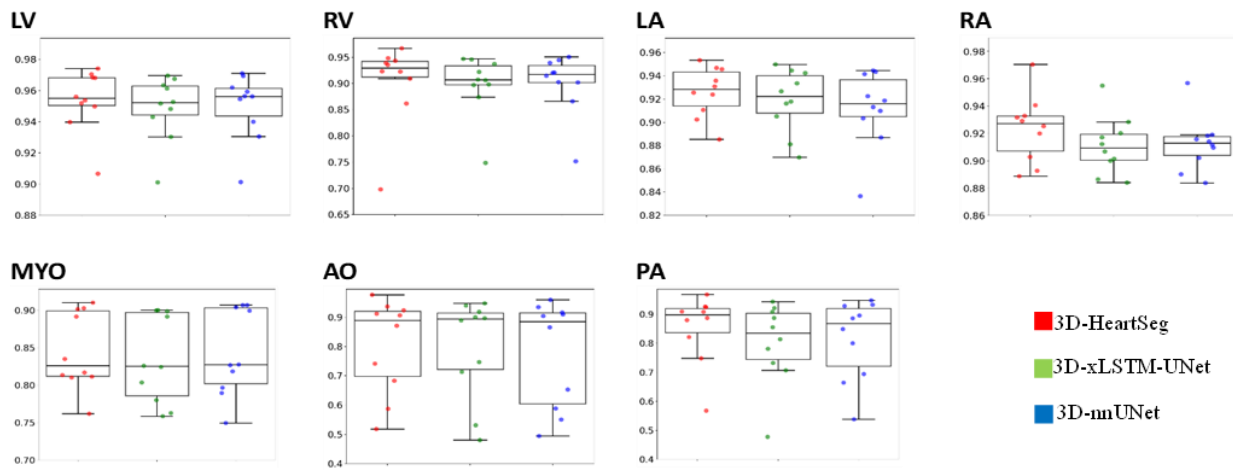

Figure 9. Performance analysis of each class in MMMRI++ dataset.

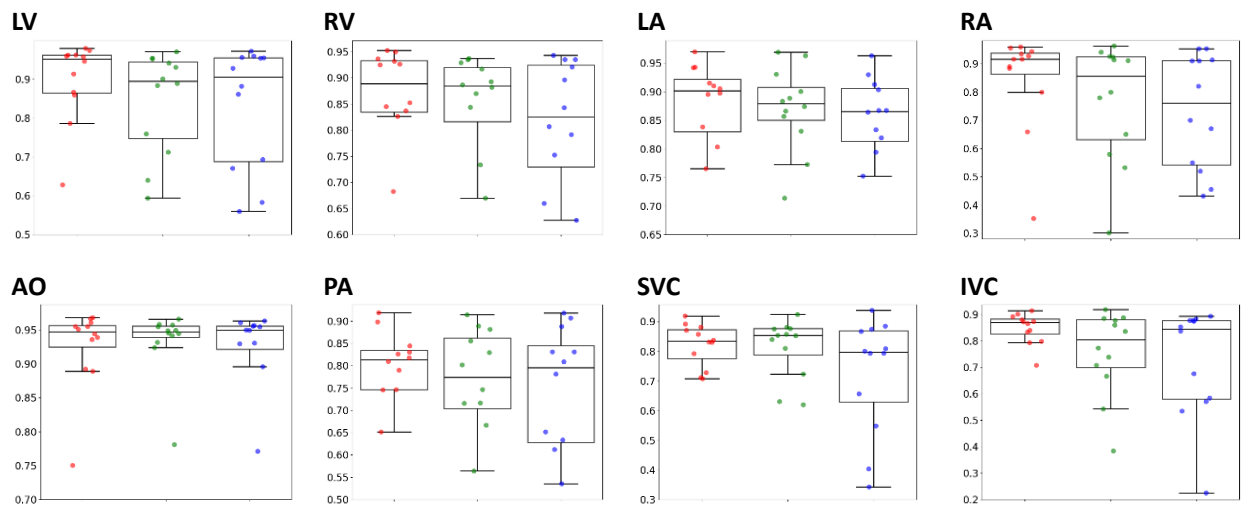

Figure 10. Performance analysis of each class in HVSMR-2.0 dataset. Red:3D-HeartSeg

, Blue:3D-nnUNet, Green:3D-xLSTM-UNet.

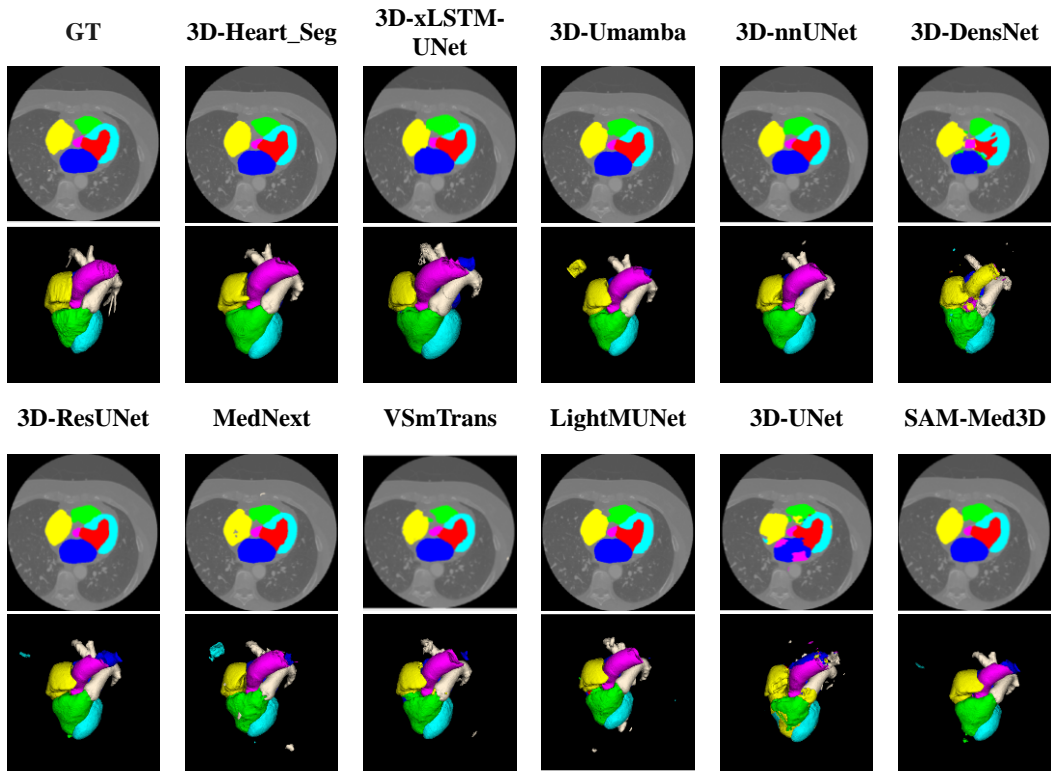

Figure 11. Performance comparison of the proposed model and existing state-of-the-art (SOTA) models using Case2006 from the MMCT test data.

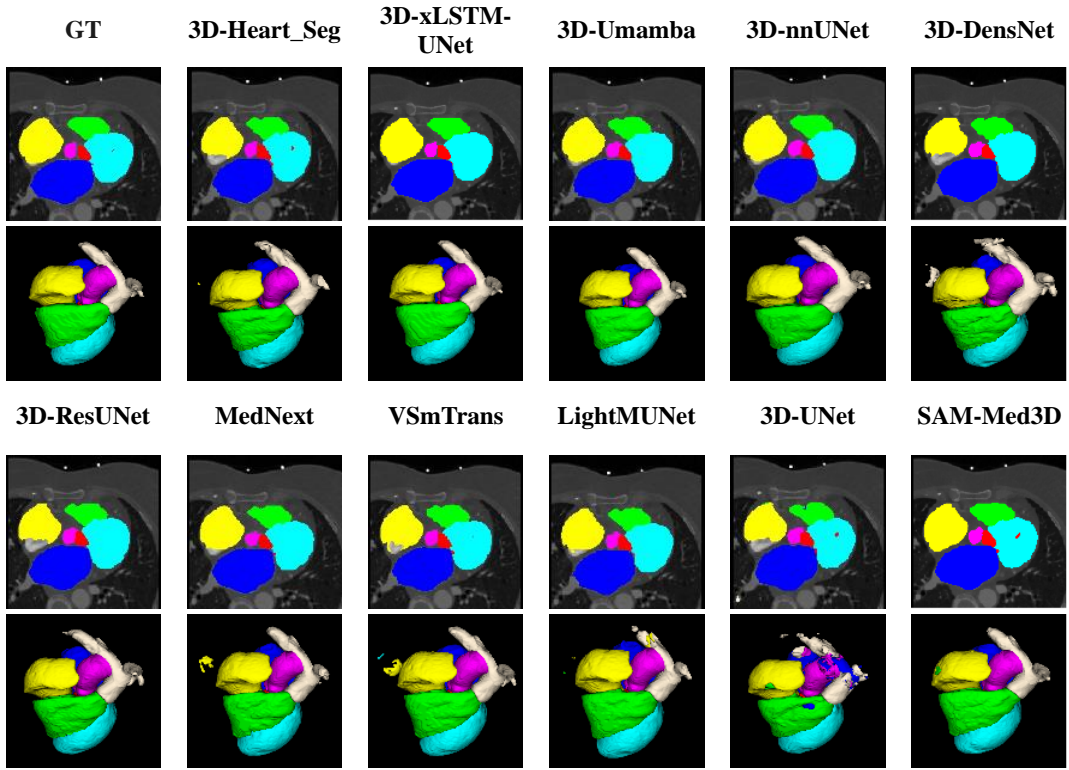

Figure 12. Performance comparison of the proposed model and existing state-of-the-art (SOTA) models using Case2020 from the MMCT++ test data.

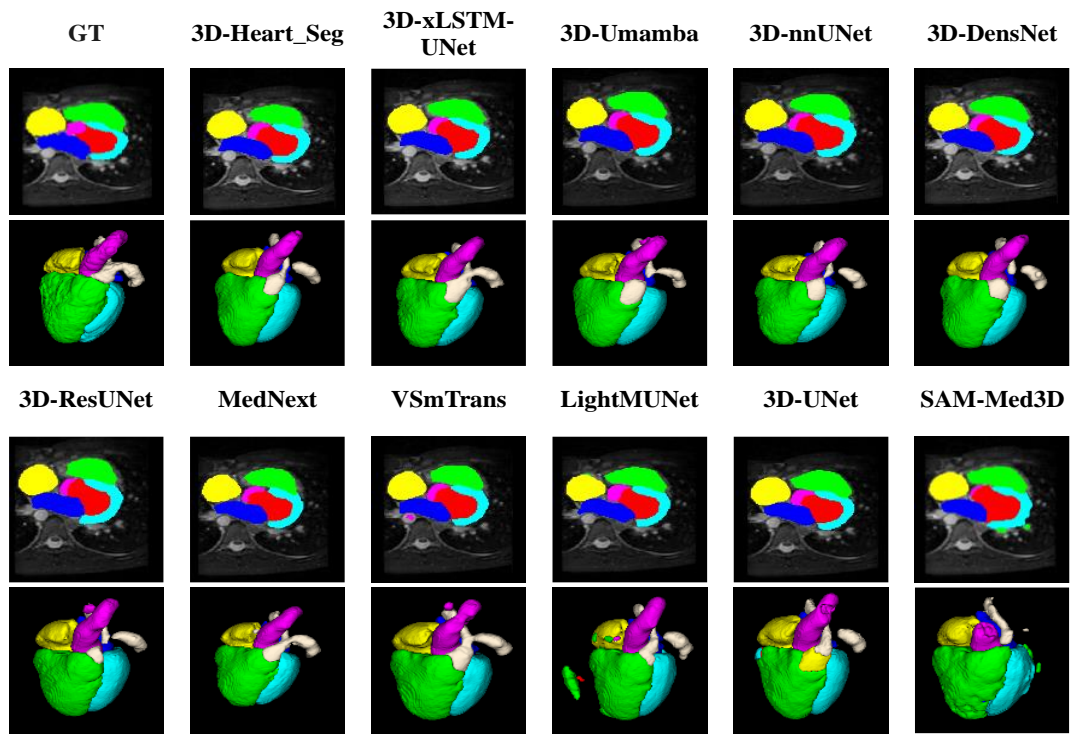

Figure 13 Performance comparison of the proposed model and existing state-of-the-art (SOTA) models using case2006 from MMMRI test data.

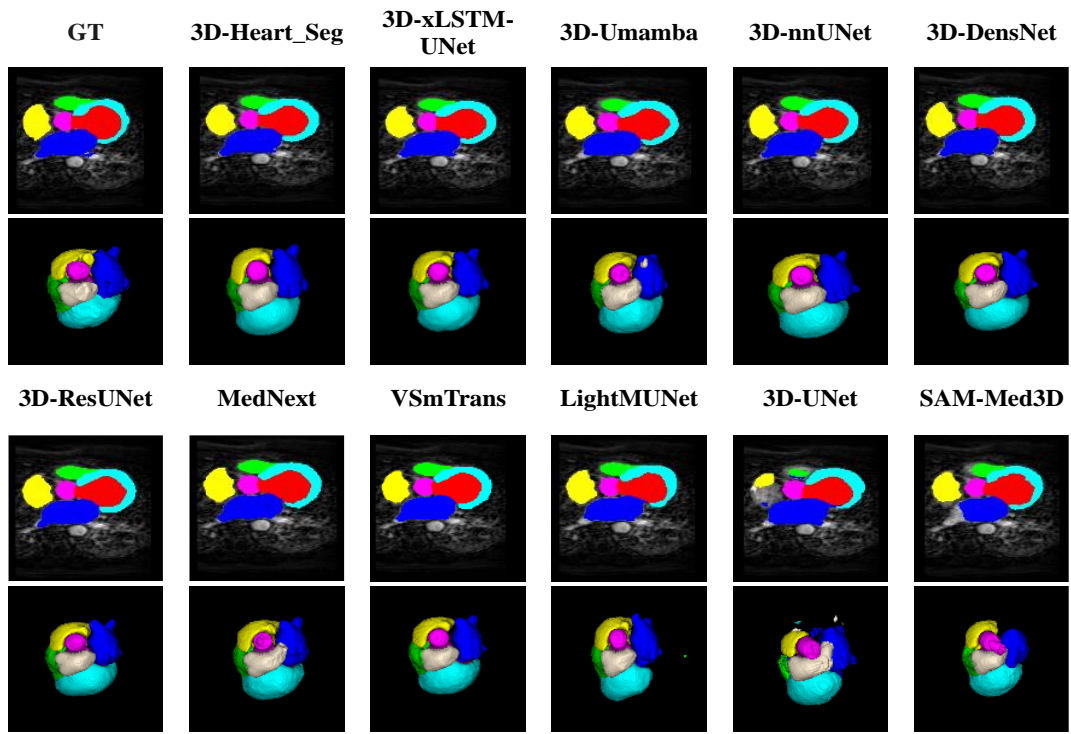

Figure.14. Performance comparison of the proposed model and existing state-of-the-art (SOTA) models using Case5024 from the MMMRI++ test data.

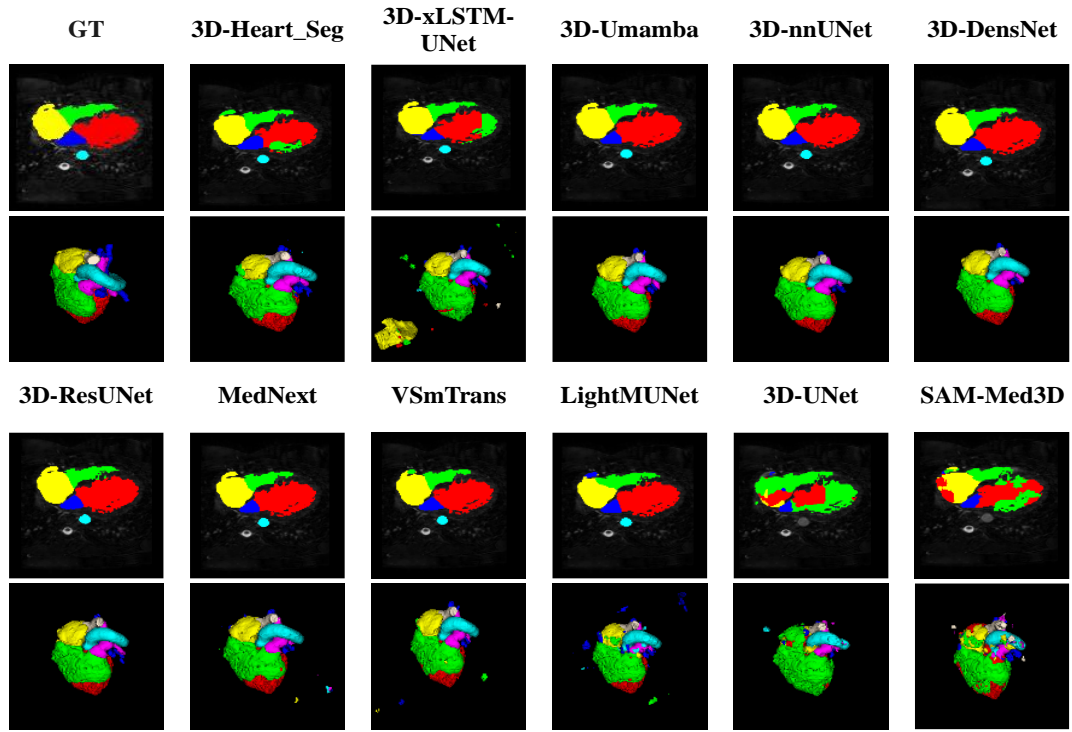

Figure. 15. Performance comparison of the proposed model and existing state-of-the-art (SOTA) models using Case009 from the HVSMR-v2 test data.

When evaluated on the MMMRI and MMMRI++ datasets, Heart_Seg3D maintains its superior performance, demonstrating consistent segmentation quality across multi-modal MRI datasets. Notably, the model exhibits enhanced adaptability to different MRI sequences, effectively mitigating modality-specific biases that often degrade segmentation accuracy in traditional approaches.

One of the key advantages of Heart_Seg3D is its ability to balance local segmentation detail and global anatomical context. The xLSTM backbone allows the model to capture long-range spatial dependencies, ensuring that even small, intricate structures like the pulmonary artery and aorta are accurately segmented while maintaining overall anatomical consistency. This is particularly important for clinical applications where fine-grained segmentation details directly impact diagnosis and treatment planning.

The visualization of segmentation mask overlays on 2D axial images demonstrates that 3D-Heart-Seg produces segmentation results that are more closely aligned with the ground truth (GT) compared to other models is shown in Figure 10-14. This highlights its ability to accurately capture anatomical structures with minimal deviation.

Further analysis of 3D segmentation masks reinforces these findings. While 3D-Heart-Seg maintains structural integrity and accurately delineates boundaries in the 3D space, several competing models exhibit segmentation errors. For instance, 3D-Umamba introduces noticeable errors in 3D segmentation, where misclassified regions and boundary artifacts can be observed in Figure.10. Similarly, models such as MedNext and LightMUNet show inconsistencies in their 3D segmentation maps, leading to visible distortions and missing anatomical details. These errors suggest a lack of robustness in handling complex structures, likely due to architectural limitations or suboptimal learning of spatial features.

Overall, our analysis confirms that 3D-Heart-Seg outperforms existing SOTA models providing a more reliable and precise representation of anatomical structures. The superior performance of our model can be attributed to its advanced feature extraction, effective spatial encoding, and improved generalization across different data variations.

The scatter plots shown in Figure.15 visually compare the predicted heart volumes from our foundation model against the ground truth across multiple datasets. Ideally, all points would lie along with the perfect agreement line (black dashed line), indicating a perfect match between predictions and ground truth. While the foundation model demonstrates generally good alignment with this line, some datasets, particularly MMCT and MMMRI, exhibit more variability, suggesting inconsistencies or potential limitations in the model's predictions for certain data subsets. In some cases, the model appears to overestimate or underestimate heart volumes in specific ranges, which could indicate biases inherent to the predictions. Overall, although our foundation model performs well across these datasets, there is an opportunity to refine its generalization and reduce prediction errors, especially for extreme heart volumes. Addressing these areas would enhance the model's accuracy and reliability, leading to more robust predictions across diverse cardiac datasets.

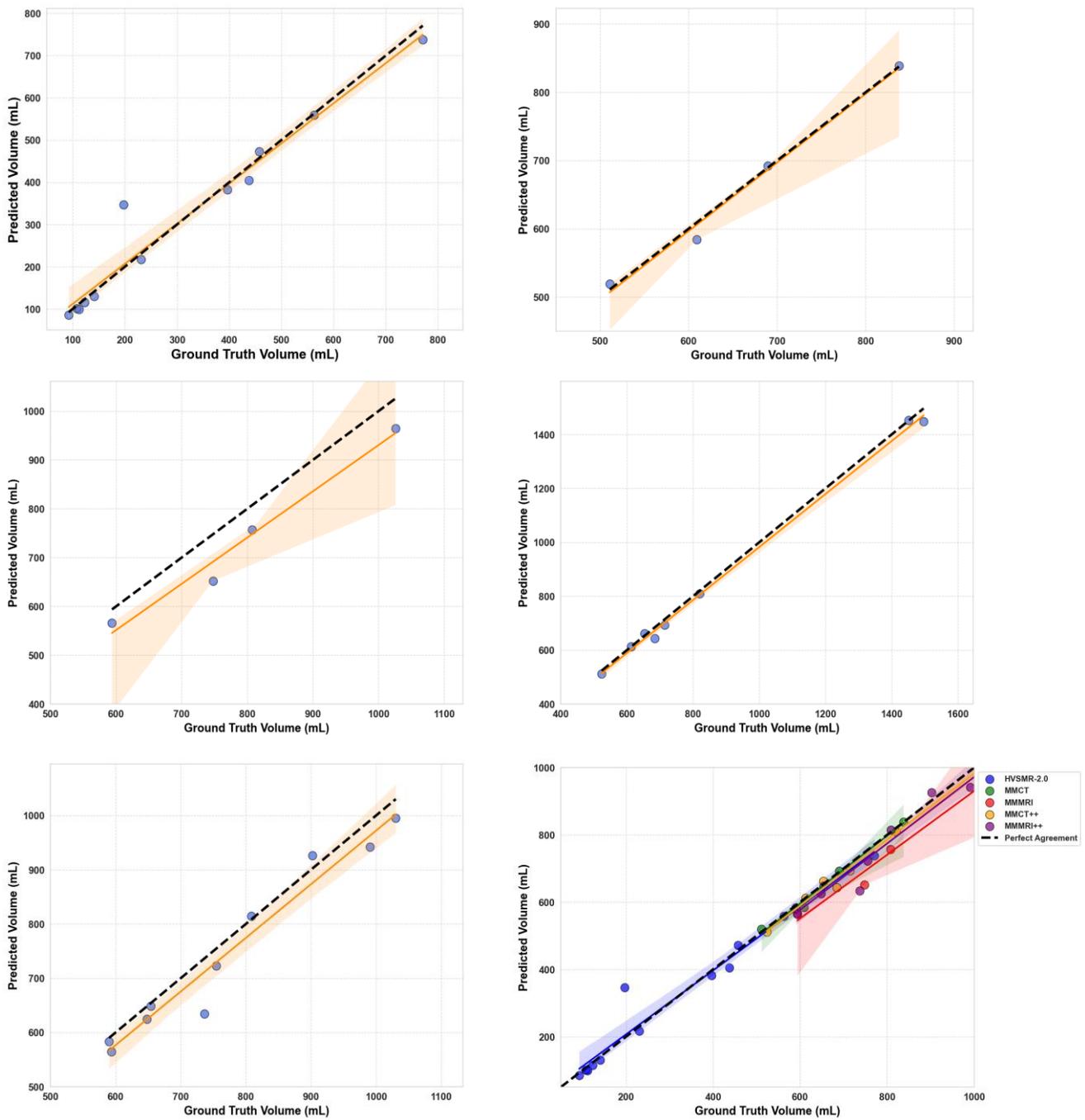

Figure.16. The scatter plots comparing the ground truth and predicted heart volumes using the foundation model across different datasets. Each subplot corresponds to a specific dataset: (a) HVSMR-v2, (b) MMCT, (c) MMMRI, (d) MMCT++, (e) MMMRI++, and (f) a combined analysis of all datasets. The black dashed line represents perfect agreement, where predicted values match ground truth volumes exactly. While the foundation model performs well across datasets, deviations from the agreement line indicate prediction errors, with certain datasets showing greater variability. The combined dataset analysis (f) provides insights into overall model performance and generalizability across different imaging modalities.

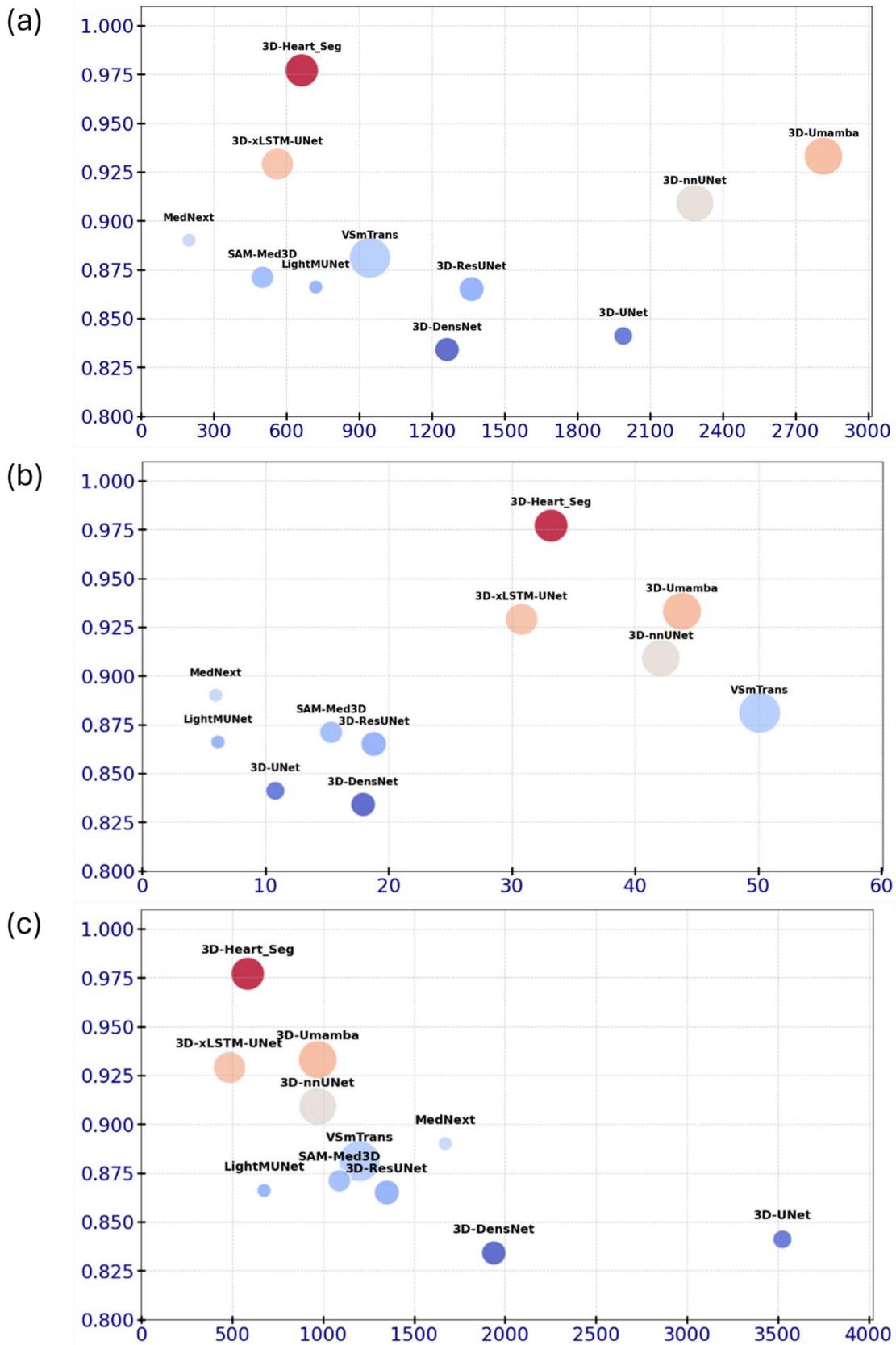

Figure 17. Computational performance comparison of the proposed model and state-of-the-art (SOTA) models: (a) Accuracy vs. Total FLOPs, (b) Accuracy vs. Total Parameters, and (c) Accuracy vs. Activation Memory Size.

Figure.17(a) presents a bubble chart illustrating the trade-off between accuracy and total FLOPs for various 3D medical image segmentation models. The x-axis represents the total FLOPs, which indicates the computational cost of each model, while the y-axis denotes accuracy, reflecting the segmentation performance. Each bubble's size likely corresponds to another factor, such as the number of parameters or memory usage, while the color may indicate different performance characteristics. Notably, 3D-Heart_Seg achieves the highest accuracy (0.98) with a moderate FLOPs count (600), making it one of the most efficient models. Meanwhile, 3D-Umamba also demonstrates high accuracy (0.95) but at a significantly higher computational cost (2700 FLOPs). Other models, such as 3D-xLSTM-UNet and 3D-nnUNet, perform well in accuracy but vary in FLOPs efficiency. On the other hand, models like 3D-UNet and 3D-DenseNet show relatively lower accuracy (0.82-0.86) despite having high FLOPs, indicating less efficiency. This visualization highlights the balance between computational cost and segmentation performance, helping in selecting models based on the desired trade-off between accuracy and efficiency.

Figure.17(b) is a bubble chart illustrating the relationship between accuracy (y-axis) and total parameters (x-axis) for various 3D medical image segmentation models. 3D-Heart_Seg achieves the highest accuracy (0.98) with a moderate parameter count (30M), suggesting an efficient balance between accuracy and model complexity. 3D-Umamba and 3D-nnUNet also achieve high accuracy (0.93-0.95) but require a significantly larger number of parameters (40M). On the other hand, models such as 3D-UNet and 3D-DenseNet have lower accuracy (0.82-0.85) despite having a moderate number of parameters, indicating they may be less efficient. VSmTrans has a large parameter count (50M) but does not achieve the highest accuracy, which may suggest diminishing returns in increasing model size. This visualization highlights the trade-off between model size and segmentation accuracy, aiding in the selection of models based on parameter efficiency and performance. The

Figure.17(c) is a bubble chart depicting the relationship between accuracy (y-axis) and activation memory size (x-axis) for various 3D medical image segmentation models. The x-axis represents the activation memory size, which indicates the computational and memory demands of each model, while the y-axis denotes accuracy, reflecting segmentation performance. 3D-Heart_Seg achieves the highest accuracy (0.98) while maintaining a moderate activation memory requirement (600MB), making it highly efficient. 3D-Umamba and 3D-xLSTM-UNet also attain high accuracy (0.93-0.95) but with relatively larger activation memory sizes (800MB-1000MB). In contrast, 3D-UNet and 3D-DenseNet require significantly more memory (2000MB-3500MB) while achieving lower accuracy (0.82-0.85), indicating inefficiency in memory usage. VSmTrans, with an activation memory size of around 1100MB, achieves moderate accuracy (0.88), but its bubble size suggests a higher computational cost. This visualization effectively highlights the trade-off between accuracy and memory efficiency, aiding in selecting models based on available computational resources and segmentation performance.

Figure 18 emphasizes the radar plot to assess the model's ability to generalize across imaging modalities, consistently outperforming SOTA models in both CT and MRI datasets for whole-heart segmentation. Through SSL pre-training, Heart_seg3d learns modality-independent features, making it highly adaptable to clinical environments where multimodal imaging is a necessity. This cross-modality capability enhances its versatility, ensuring reliable performance across diverse clinical datasets. The radar plot in Figure 16 provides a comparative analysis of the Dice scores achieved by our proposed 3D-Heart_Seg foundation model and various state-of-the-art (SOTA) models, including 3D-xLSTM-UNet, 3D-Umamba, 3D-nnUNet, 3D-DenseNet, 3D-ResUNet, MedNext, VSmTrans, LightMUNet, 3D-UNet, and SAM-Med3D. The performance is evaluated for both CT and MRI modalities, illustrating the segmentation accuracy across different architectures.

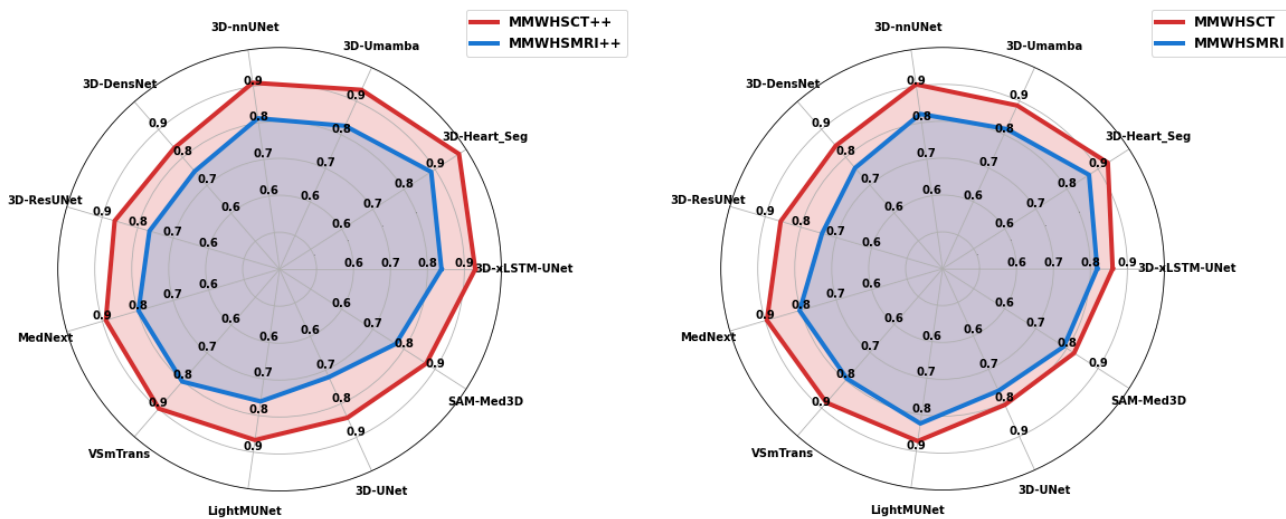

Figure 18. Radar plot of MRI vs CT Comparison of Dice Score using proposed and SOTA models.

## Discussion

We proposed a 3D student-teacher model trained on a large unlabeled medical dataset, inspired by the 2D DINOv2 vision model. Whole-heart segmentation from CT and MRI scans is a crucial step in cardiovascular disease analysis, as it enables detailed anatomical assessment and facilitates downstream clinical applications such as diagnosis, treatment planning, and surgical interventions. Despite significant advancements in medical image segmentation, existing methods face challenges due to modality-specific biases and the need for extensive labeled datasets. CT and MRI scans have inherent differences in contrast, resolution, and tissue representation, making it difficult for traditional models to generalize across both modalities. Additionally, the requirement for manually labeled datasets increases the burden on clinicians and restricts the scalability of deep learning-based segmentation approaches. Addressing these challenges necessitates a robust framework that can leverage large amounts of unlabeled data while ensuring strong generalization capabilities across multiple imaging modalities.

To overcome these limitations, we propose a foundation model for whole-heart segmentation based on a 3D self-supervised learning (SSL) framework that employs a student-teacher architecture. Self-supervised learning has shown promising results in various domains by enabling models to learn meaningful representations without the need for extensive labeled data. Our approach involves pretraining the model on a large, unlabeled dataset of CT and MRI scans, allowing it to learn shared anatomical features and modality-independent structures. The student-teacher architecture facilitates knowledge distillation, where the teacher network guides the student model to improve its representations, ultimately enhancing segmentation performance. By eliminating the reliance on fully supervised training, our method significantly reduces the dependency on manual annotations, making it more efficient and scalable for clinical applications.

Our model is the incorporation of the xLSTM backbone, which is specifically designed to capture long-range spatial dependencies and complex anatomical structures in 3D medical images. Unlike conventional convolutional neural networks (CNNs) that primarily rely on local spatial features, xLSTM effectively models global dependencies, which is essential for accurate whole-heart segmentation. The integration of xLSTM enhances the model's ability to recognize intricate heart structures, improving segmentation accuracy and consistency across varying imaging conditions. Furthermore, the model's multi-modal pretraining ensures robust generalization to both CT and MRI scans, mitigating modality-specific variations and allowing seamless adaptation to diverse clinical settings.

We utilized approximately 49,048 unlabeled datasets in our SSL framework. Initially, we performed SSL on 14,048 CMR short-axis images. We then conducted reSSL on 35,000 CT and MRI whole 3D volumes, further refining our model's understanding of multi-modal anatomical structures. This extensive pretraining allowed our model to develop strong feature representations and improved its generalization ability across different imaging modalities.

To validate our approach, we introduce xLSTM-UNet-based architecture for downstream whole-heart segmentation tasks. UNet has been widely recognized for its effectiveness in medical image segmentation due to its encoder-decoder structure, which captures both high-level contextual information and fine-grained spatial details. By integrating xLSTM with UNet, we enhance the model's ability to handle complex anatomical variations while maintaining high segmentation fidelity. The proposed architecture is evaluated on few-label CT and MRI datasets, demonstrating its ability to perform accurate segmentation with minimal labeled data. Our results indicate that the model outperforms conventional methods in both segmentation accuracy and generalization capability, highlighting its potential for clinical implementation.

The implications of our research are significant for the medical imaging community, as our foundation model provides a scalable and efficient solution for whole-heart segmentation. By leveraging large-scale unlabeled datasets and a self-supervised learning paradigm, our approach minimizes the challenges associated with data annotation while ensuring high-performance segmentation across different modalities. Future work can explore further enhancements, such as incorporating additional imaging modalities, refining the model's architecture, and integrating domain adaptation techniques to improve performance on unseen datasets. Overall, our study contributes to the advancement of automated whole-heart segmentation, paving the way for more precise and reliable cardiovascular disease analysis in clinical practice.

## Conclusion

Our findings validate that Heart_seg3d is a robust and highly effective foundation model for whole-heart segmentation. By integrating self-supervised learning, leveraging multi-layer SSL pre-training, and implementing a hierarchical feature extraction mechanism, the model achieves state-of-the-art performance across both CT and MRI datasets. The ability to generalize across imaging modalities ensures its suitability for clinical applications, reinforcing its potential as a leading solution in cardiac imaging segmentation tasks. Future work will focus on expanding the model's adaptability to other anatomical structures and further optimizing computational efficiency for real-time clinical deployment.

## Acknowledgment

This work is supported by the Technology Missions Fund under the EPSRC Grant EP/X03870X/1, the British Heart Foundation (RG/20/4/34 803), and The Alan Turing Institute, London, UK.